\documentstyle[epsf,longtable]{myl-aa}

\begin {document}
   \thesaurus{13 
              (11.02.2; 
               13.07.2) 
              }
   \title{The temporal characteristics of the TeV Gamma emission
from Mkn 501 in 1997 - Part II: Results from HEGRA CT1 and CT2}

\author
{F. Aharonian\inst{1},
A.G. Akhperjanian\inst{7},
J.A.~Barrio\inst{2,3},
K.~Bernl\"ohr\inst{1,8},
H. Bojahr\inst{6},
J.L. Contreras\inst{3},
J. Cortina\inst{3,2},
A. Daum\inst{1},
T. Deckers\inst{5},
S. Denninghoff\inst{2},
V. Fonseca\inst{3},
J. Gebauer\inst{2},
J.C. Gonzalez\inst{3},
G. Heinzelmann\inst{4},
M. Hemberger\inst{1},
G. Hermann\inst{1},
M. Hess\inst{1},
A. Heusler\inst{1},
W. Hofmann\inst{1},
H. Hohl\inst{6},
D. Horns\inst{4},
A. Ibarra\inst{3},
R. Kankanyan\inst{1},
M. Kestel\inst{2},
O. Kirstein\inst{5},
C. K\"ohler\inst{1},
A. Konopelko\inst{1},
H. Kornmayer\inst{2},
D. Kranich\inst{2},
H. Krawczynski\inst{1,4},
H. Lampeitl\inst{1},
A. Lindner\inst{4},
E. Lorenz\inst{2},
N. Magnussen\inst{6},
H. Meyer\inst{6},
R. Mirzoyan\inst{2},
A. Moralejo\inst{3},
L. Padilla\inst{3},
M. Panter\inst{1},
D. Petry\inst{2,6,9},
R. Plaga\inst{2},
A. Plyasheshnikov\inst{1},
J. Prahl\inst{4},
G. P\"uhlhofer\inst{1},
G. Rauterberg\inst{5},
C. Renault\inst{1},
W. Rhode\inst{6},
A. R\"ohring\inst{4},
V. Sahakian\inst{7},
M. Samorski\inst{5},
D. Schmele\inst{4},
F. Schr\"oder\inst{6},
W. Stamm\inst{5},
B. Wiebel-Sooth\inst{6},
C. Wiedner\inst{1},
M. Willmer\inst{5},
H. Wirth\inst{1},
W. Wittek\inst{2}}
\institute{Max-Planck-Institut f\"ur Kernphysik,
Postfach 103980, D-69029 Heidelberg, Germany \and
Max-Planck-Institut f\"ur Physik, F\"ohringer Ring
6, D-80805 M\"unchen, Germany \and
Universidad Complutense, Facultad de Ciencias
F\'{\i}sicas, Ciudad Universitaria, E-28040 Madrid, Spain \and
Universit\"at Hamburg, II. Institut f\"ur
Experimentalphysik, Luruper Chaussee 149,
D-22761 Hamburg, Germany \and
Universit\"at Kiel, Institut f\"ur Kernphysik,
Leibnitzstr. 15, D-24118 Kiel, Germany \and
Universit\"at Wuppertal, Fachbereich Physik,
Gau{\ss}str.20, D-42097 Wuppertal, Germany \and
Yerevan Physics Institute, Alikhanian Br. 2, 375036 Yerevan, Armenia \and
Now at Forschungszentrum Karlsruhe, P.O. Box 3640, D-76021 Karlsruhe \and
Now at Universidad Aut\'{o}noma de Barcelona,
Institut de F\'{\i}sica d'Altes Energies, E-08193 Bellaterra, Spain}


   \date{Received ; accepted}

   \maketitle

\markboth{F. Aharonian et al.: 
Temporal characteristics of the TeV $\gamma$-emission
from Mkn 501 - Part II}{F. Aharonian et al.: 
TeV characteristics of Mkn 501, HEGRA CT1 and CT2 observations}

\begin{abstract}
We present the results on the TeV gamma emission from Mkn 501 in 1997
obtained with the stand-alone Cherenkov telescopes CT1 and CT2
(threshold $\ge 1.2$ and $\ge 1.0$ resp.) of the
HEGRA collaboration. 
The CT1 lightcurve has the most complete coverage
of all TeV observations of Mkn 501 in 1997 due to the additional
observations made under the presence of moonlight.
CT2  - at the time of these observations a second generation
Cherenkov telescope with relatively low imaging resolution - 
is a well tested instrument and its 85 hours of observational
data on Mkn 501 in 1997 are useful for consistency 
checks and provide some additional daily data points. The Mkn 501
lightcurve data show significant correlation with the RXTE 
ASM X-ray data consistent with no time-lag. The spectral analysis shows 
a steepening spectrum extending beyond 10 TeV. No change of the
spectral slope with the variation of the intensity was found.
The results presented here are consistent with the results
from the HEGRA Cherenkov telescope system presented in part I of this paper.

\end{abstract}
\section{ Introduction }
The BL-Lac object Mkn 501 showed strong and frequent
flaring in 1997. 
The source has been observed by many different experiments using
imaging air Cherenkov telescopes (IACTs).
Here we report on observations with the HEGRA stand-alone telescopes CT1 and
CT2 while observations with the HEGRA CT system are reported in part I of
this paper (Aharonian et al. \cite{aharonian98}, subsequently {\it Part I}).

From March 11 to October 20, 1997 the source was monitored every night
whenever weather and background light permitted it. A fraction
of the observations
was carried out by up to 6 telescopes.

A detailed discussion of Mkn 501 and its history of $\gamma$-emission is
given in Part I, as well as
the details of the stereo-mode observations with 4 telescopes, the
related analysis methods, the stereo-mode results and some comparisons of the
stereo-mode data with RXTE observations.
The data from the stand alone telescopes CT1 and CT2 are less precise
than the system data as energy and angular resolution are somewhat
worse. Nevertheless they complement th CT system data as significantly 
longer observations were carried out. Due to additional observations
made under the presence of moonlight the lightcurve of CT1 is the most 
complete of all observations in 1997.

This paper has the following structure: Section 2 summarizes the
relevant telescope parameters of CT1 and CT2
and important performance data. The details of the observations and data
analysis are presented in section 3 together with the combined lightcurve.
For the comparison with lower energy data from RXTE we include HEGRA
observations from 1996.
The CT2 data analysis is presented in section 4. The combined lightcurve,
specific details and conclusions are discussed in sections 5 and 6. 

\section{The HEGRA Cherenkov Telescopes CT1 and CT2}

The HEGRA collaboration is operating six imaging atmospheric
Cherenkov telescopes for Gamma Astronomy as part of its cosmic ray detector 
complex at the Observatory Roque de los Muchachos on the Canary island of 
La Palma (28.75$^\circ$ N, 17.89$^\circ$ W, 2200 m a.s.l., see e.g. Lindner 
et al. \cite{lindner97}). While the first two telescopes (CT1 and CT2) are operated 
in stand-alone mode, the other four (CT3, 4, 5 and 6) are run as a 
system of 
telescopes in order to achieve stereoscopic observations of the air-showers.

\subsection{The telescope CT1}

HEGRA CT1 was commissioned in August 1992. In its 1997 configuration, 
CT1 had a mirror made up of 18 spherical round glass mirrors of 5 m focal 
length and a total mirror area of 5 m$^2$. The photomultiplier (PM)
camera of CT1 consists of 127 3/4$''$ EMI tubes  9083A in a hexagonally 
dense package with 
an angular diameter of $\approx 3^\circ$ (individual pixel diameter:
 $0.25^\circ$).
The tracking accuracy of CT1 is better than 0.1$^\circ$. The telescope
hardware is described in detail
in Mirzoyan et al. (\cite{mirzoyan94}) and Rauterberg 
et al. (\cite{rauterberg95}).

\subsubsection{Camera settings and observations under the presence of
moonlight}
During the 1997 observing period, CT1 was run with a range of 
slightly different 
high voltage settings for the PM camera:

During dark nights, two settings were used: 
Before 29th of April the 
settings from previous years had been used which we name HV1 in this 
paper. After the 29th of April, the high voltages were increased by 
$\approx 6$\,\% in order to compensate for PM dynode aging effects and 
to lower
the energy threshold of the telescope to 
below its pre-1996 value. This second 
setting we denote by HV2.

Soon after the beginning of the 1997 observing period, the strong 
variability of Mkn 501 made it obvious that it was of great importance to 
dedicate as much observation time as possible to the source. Until 
recently, it was believed that Cherenkov telescopes can only operate 
during moonless nights due to the increase in PM current and noise caused 
by the general increase in background light. As our studies with CT1 show, 
this limitation can be largely overcome 
by fast amplifiers with AC coupling to 
 low-gain PM cameras for which
the high voltage is reduced by several percent compared to the optimal
setting for moonless nights. The used voltage reduction 
increases the telescopes
threshold by a factor up to 2.6, 
but observations of strong gamma sources can still give 
useful results.
Details of the observation in the presence of moonlight are given in
Raubenheimer et al. (\cite{raubenheimer98}).

CT1 has observed Mkn 501 for nearly 7 months whenever the weather fulfilled the 
standard observing conditions and the source was at zenith angles below
$\approx 60^\circ$. The additional
observations under the presence of moonlight make
the lightcurve obtained from CT1 the most complete one compared to 
all other 1997 light curves of this source in the TeV energy range. The 
moonlight
observations were taken with four different PM voltage settings: HV1 and 
HV2 as described above and settings with voltage reduced by 10\% and 14\% 
relative to HV2. The latter settings are named HV3 and HV4.
Nearly all the data taken under the presence of moonlight were taken
with the settings HV1 and HV2. HV4 was used only for observations close to the
nearly full moon.

\subsection{The telescope CT2}

The second HEGRA Cherenkov telescope,
CT2, was built in 1993 and has been observing in an essentially unchanged 
configuration since 1994. CT2 is located at 93\,m distance from CT1, i.e.\ 
some of the showers are seen simultaneously by both telescopes when
operated at the same time. Nevertheless, we treat the observations as
independent ones.

CT2 was the prototype for the
HEGRA Cherenkov telescope system. As opposed to the equatorially mounted CT1,
it has an ALT-AZ mount. The mirror elements 
are again round glass mirrors of 60 cm $\oslash$ and 5 m focal length,
but 30 instead of 18 are used which give CT2 a mirror
area of 8.5 m$^2$ and thus a lower energy threshold compared to CT1.

In 1997 CT2 was still operated with its original 61 pixel camera with a field of view
of 3.7$^\circ$ and an angular diameter of the individual pixel of 0.43$^\circ$.
Studies of the trigger rate as a function of trigger threshold 
showed that the
performance of the telescope has not noticeably 
changed since 1995 and that the nominal
energy threshold of 1 TeV for primary gammas is still valid. The telescope is
described in Wiedner (\cite{wiedner94}) and Petry (\cite{petry97b}).

\subsection{Performance of the telescopes}
\label{sec_perf_of_ct}
Table \ref{param-tab} summarizes some essential 
parameters of the telescopes. Most of the
parameters were determined experimentally while some were calculated from
Monte Carlo (MC) simulations.
For the MC simulations we used the computational code developed by Konopelko (\cite{konopelko96}).
This program includes the losses of
Cherenkov light due to atmospheric effects, i.e.\ Rayleigh and 
Mie scattering, as
well as the telescope parameters such as spectral mirror reflectivity,
PM quantum efficiency etc. 
The simulations
took into account the imperfections of the telescope optics and the differences
in the CT1 PM 
noise for the different night sky background (NSB) conditions, 
e.g.\ due to the presence of moonlight. 
The relation between
photoelectrons and measured quantities, i.e., the ADC conversion factor,
has been determined by a separate experiment in 1995/96 for the HV setting
HV1, i.e.\ before the dynode aging.
For the other HV settings
of CT1 the related change in conversion factors has been calculated from
the HV-gain characteristics of the PMs, which were found to be in excellent
agreement with the change of trigger rate (after 
subtracting noise triggers). 

\begin{table*}[h]
\begin{tabular}{llll}
\hline
&CT1&CT2&comments\\
\hline
mirror area& 5 m$^2$& 8.4 m$^2$&mean reflectivity $\approx 80\%$\\
camera diameter& 3$^\circ$&3.7$^\circ$&$\approx 98\%$ active area\\
pixel no. and size&127; 0.24$^\circ$& 61; 0.43$^\circ$&\\
tracking precision&$< 0.1^\circ$&$< 0.1^\circ$&\\
trigger condition
& $\geq$ 2 coinciding pixels& $\geq$ 2 coinciding pixels&\\
&each above 12 pe (HV2)&each above 34 pe &\\
trigger rate &1.75 Hz (HV2)&2.6 Hz&raw rate in zenith position\\
trigger rate after filter&1.2 Hz (HV2)&1.4 Hz&rate after filter cuts, in
zenith position\\
threshold in zenith&HV1: 1.8 TeV&1 TeV&\\
&HV2: 1.2 TeV&&\\
&HV3: 2.4 TeV&&\\
&HV4: 3.2 TeV&&\\
angular resolution&0.13$^\circ$&0.2$^\circ$&for single events, mean of axis
of error ellipse\\
collection area&see Fig.~\ref{fig-CT1-CT2}a-c&
see Fig.~\ref{fig-CT1-CT2}d&\\
energy resolution&see Fig.~\ref{fig-eresct1}b
&see Fig.~\ref{fig-eresct1}b&\\
$Q (\gamma/h)$ after trigger& 13& 7.5&
refers to entire FOV of camera\\[-0.5ex]
and filter$^{1)}$&&&\\
$Q (\gamma/h)$$^{1)}$ &4.2 ($\alpha$ bin 0-10$^\circ$)& 
3.7 ($\alpha$ bin 0-15$^\circ$)&refers to restricted ALPHA bin\\
rate for Crab (zenith)& 10 over bg. of 6 (HV1)& 14 over bg. of 47 &\\
$\gamma/h$ cuts&dynamic supercuts&static supercuts&\\
ALPHA cut& $\leq 10^\circ$& $\leq 15^\circ$&for optimal signal/background\\
\hline
\end{tabular}
\par\smallskip
{\small 1) not including $\gamma$ enrichment due to trigger}\\
\caption{Main parameters of the HEGRA telescopes CT1 (1997 configuration)
  and CT2.}
\label{param-tab}
\end{table*}

\begin{figure*}
\vspace{-0.3cm}
\begin{center}
\begin{minipage}[b]{12.5cm}
\epsfxsize=12.5cm
\epsffile{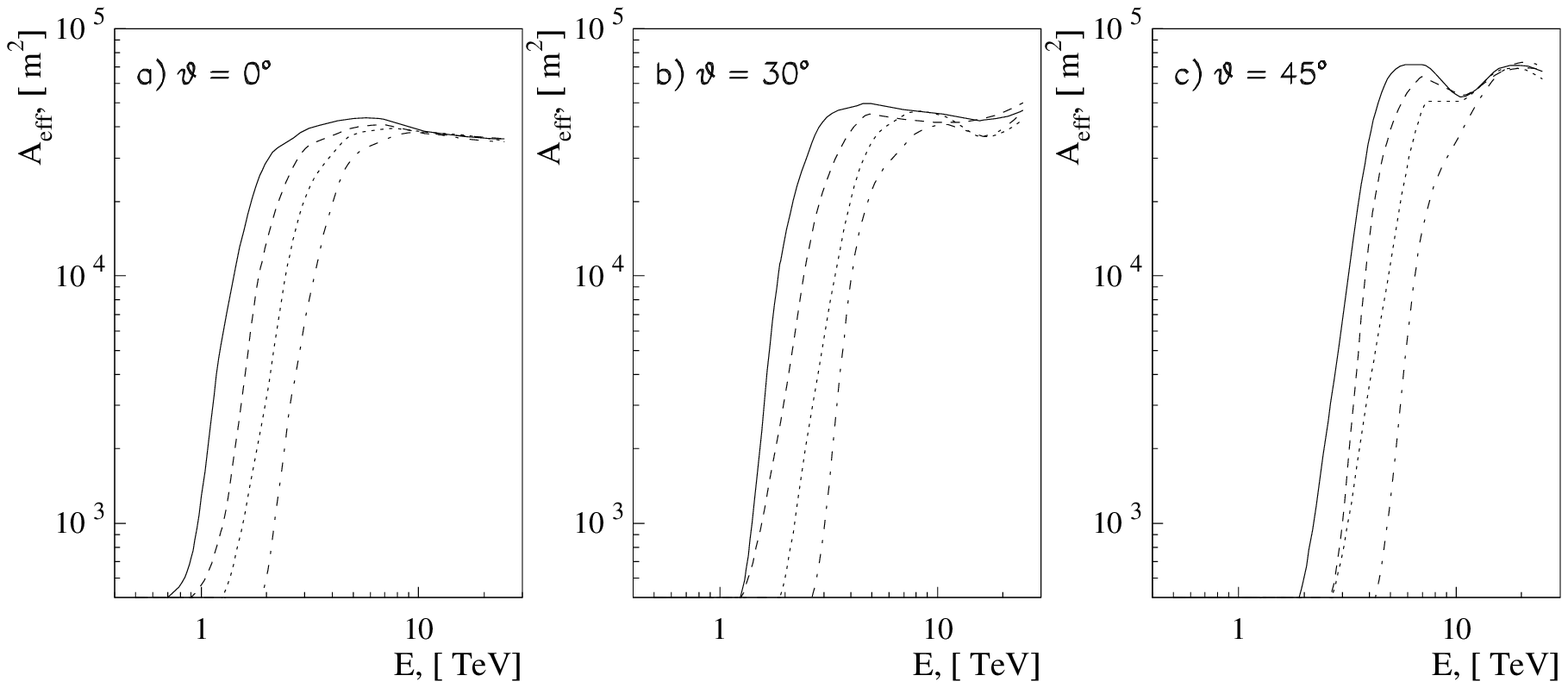}
\end{minipage}
\begin{minipage}[b]{5cm}
\epsfxsize=5cm
\epsffile{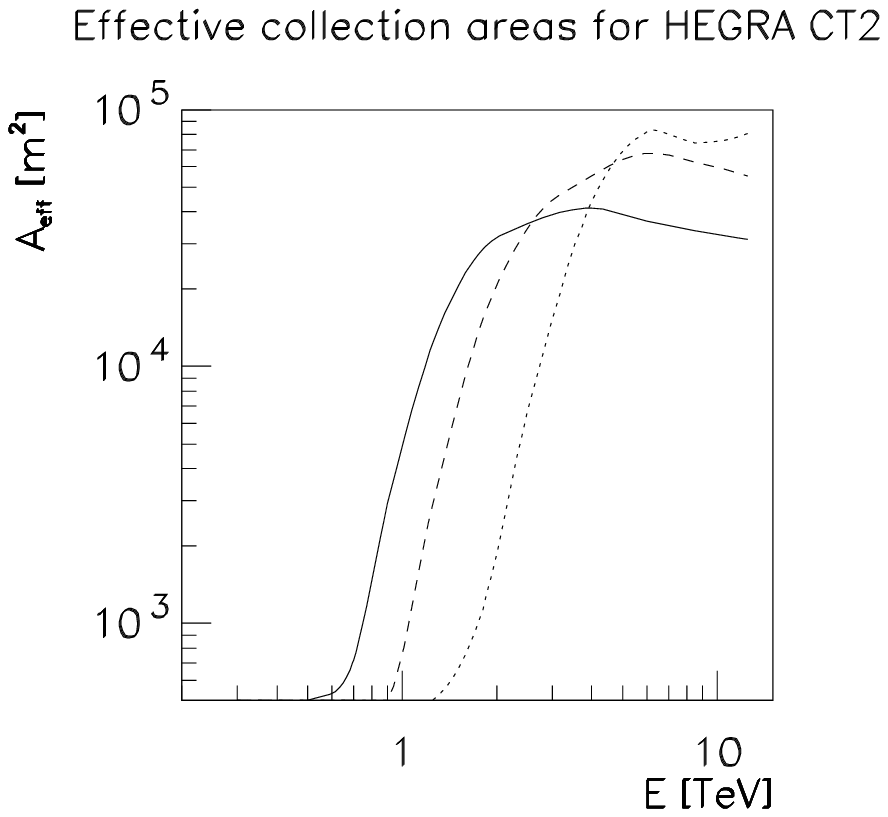}
\end{minipage}
\end{center}
\vspace{-0.5cm}
\caption{\label{fig-CT1-CT2}
(a), (b), (c) The effective collection area of CT1 (after gamma/hadron 
separation cuts) for primary gammas as a function of primary
energy for three different zenith angles. The individual 
lines denote the high voltage
settings of the camera:  solid line = HV2, dashed line = HV3,
dotted line = HV1,
dot-dashed line = HV4.
The fourth plot shows the effective collection areas of HEGRA CT2 for primary gammas
        after gamma hadron separation cuts for three 
different zenith angles:
        0$^\circ$ (solid line), 30$^\circ$ 
(dashed line), 45$^\circ$ (dotted line).}
\end{figure*}

The effective collection area depends on the HV setting, zenith angle and
used $\gamma$/hadron separation cuts. Fig.~\ref{fig-CT1-CT2}a-c shows 
the collection area
of CT1 for the four HV settings and three different zenith angles and 
Fig.~\ref{fig-CT1-CT2}d the areas for CT2, respectively.
The image cut procedures are different for CT1 and CT2. For the CT1 data
the so-called dynamical supercuts, depending on the zenith angle, the image
size and the distance parameters were used, see 
Petry \& Kranich (\cite{petrykranich97}) and Kranich (\cite{kranich97b}) for details. 
If the HV setting was the same
for moon and non-moon
observations, then 
the difference in NSB only changes the effective collection area at the
 $< 5$ \% level. This change was taken into account in flux (resp. spectrum) calculations
but is too small to be visible in Fig. \ref{fig-CT1-CT2}a-c.
Standard supercuts were used for CT2 
as in  Petry et al. (\cite{petry96}) due to the coarse pixel structure of the
camera.

\begin{figure}
\begin{minipage}[b]{8cm}
\epsfxsize=8cm
\epsfysize=6.5cm
\epsffile{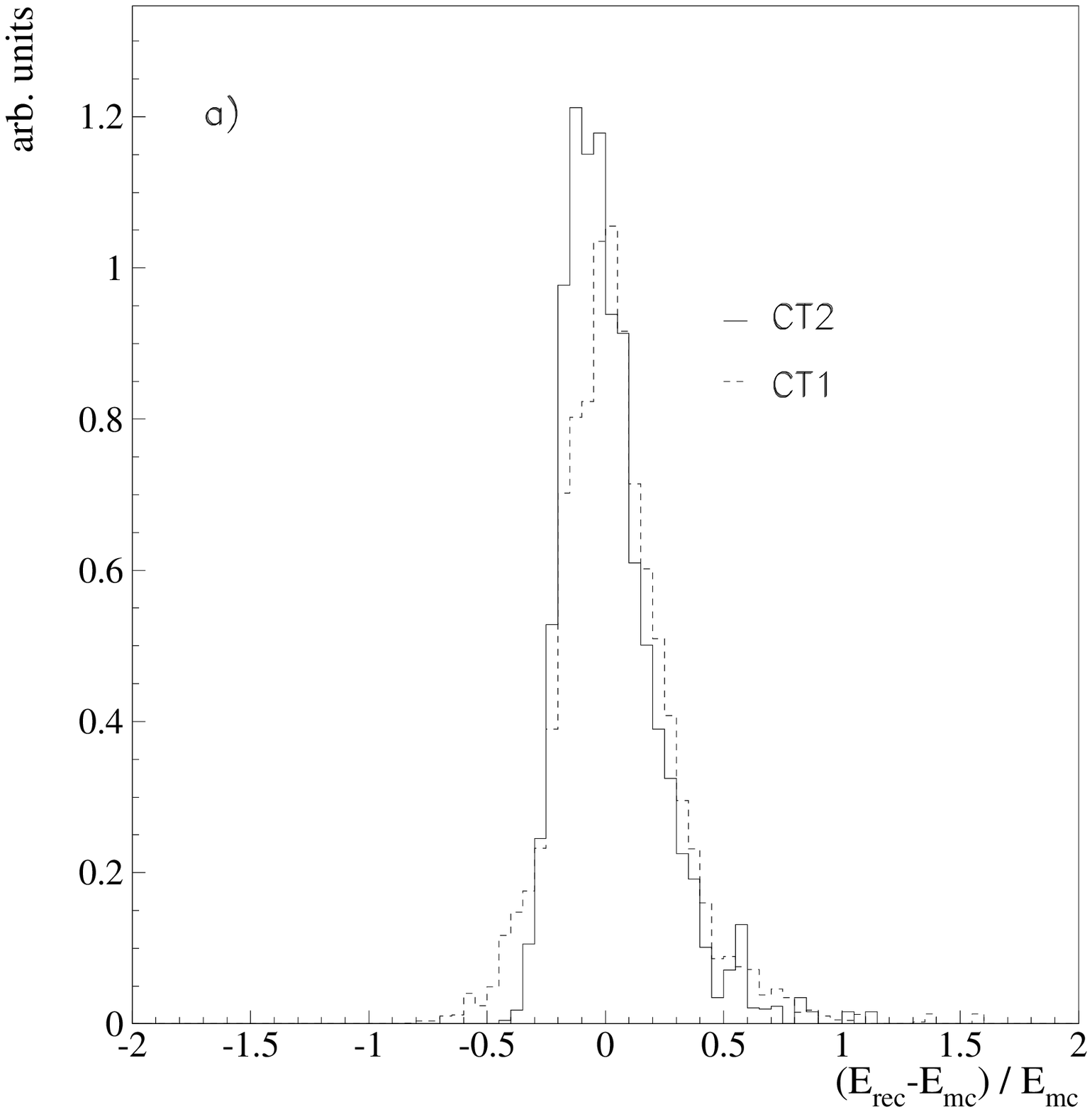}
\end{minipage}
\begin{minipage}[b]{8cm}
\epsfxsize=8cm
\epsfysize=6.5cm
\epsffile{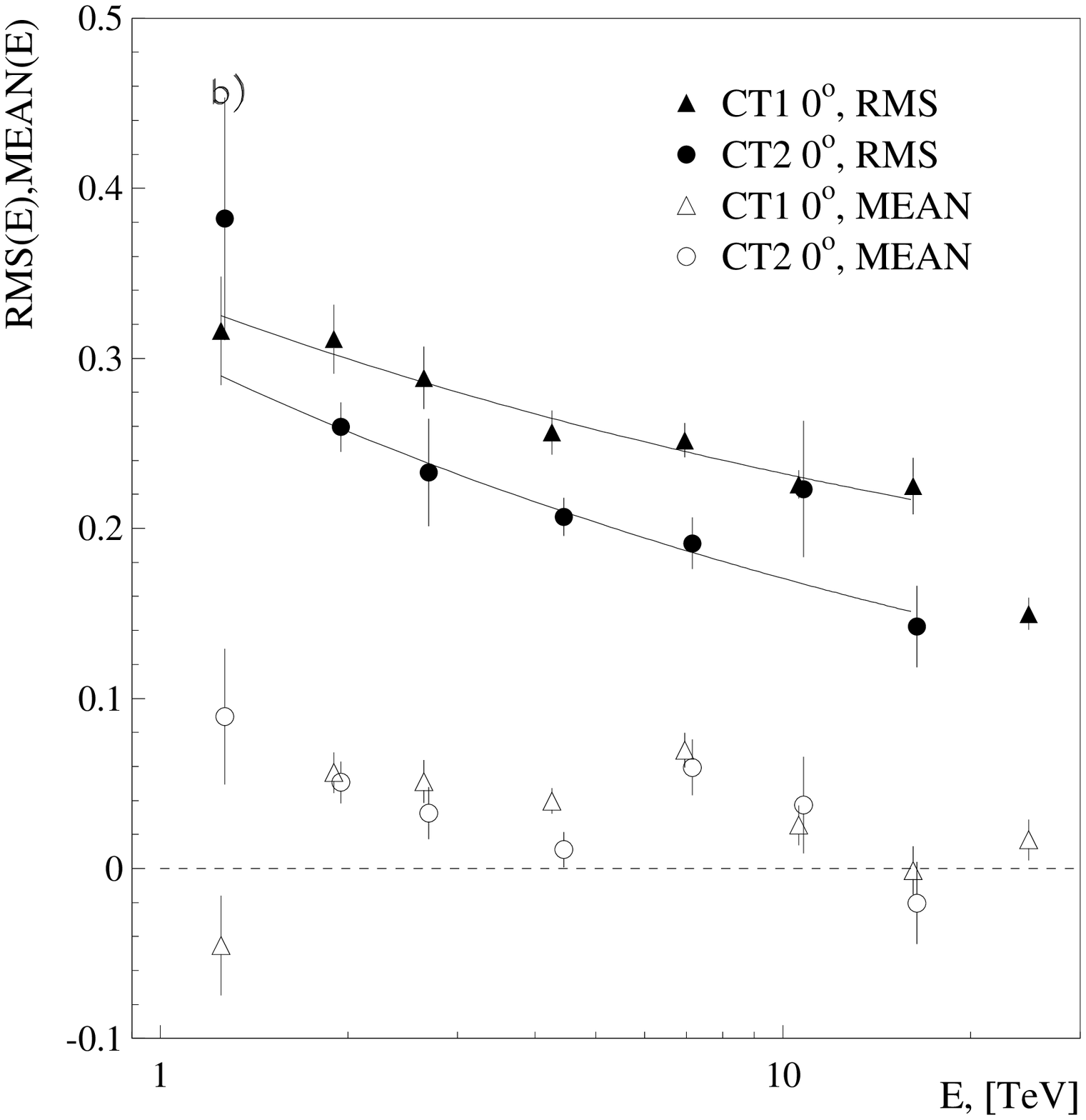}
\end{minipage}
\caption{\label{fig-eresct1}
(a) Comparison between MC input energy and reconstructed energy. For
the energy spectrum a} power law with a differential coefficient
$-2.2$ is assumed; 
(b) Predicted energy
resolution as a function of $\gamma$ energy and relative deviation between the mean
reconstructed energy and the initial energy. For CT1 and CT2.
\end{figure}

The energy reconstruction as well as the energy resolution of both
telescopes depend mainly on the image parameter SIZE. The SIZE is
in first order a good approximation of the initial $\gamma$ energy. In
second order one has to apply corrections due to the zenith angle and
the impact parameter. Also intrinsic fluctuations in the height of the
shower maximum, xmax, can affect the energy reconstruction. With a
single telescope one cannot determine the impact parameter
directly. Nevertheless, the image parameter DIST provides a
sufficiently precise measure of the impact parameter, while up to now
no equivalent observable for xmax is known. From MC simulations, as
well as from accelerator experiments it is known, that electromagnetic
showers have a much smaller fluctuation of the depth of the shower maximum compared to hadronic
showers. From MC data we developed a correction function which allowed
us to calculate the initial energy, as well as predicting the energy
resolution from the image parameters SIZE, DIST, WIDTH and the zenith
angle. We used the Levenberg-Marquardt method(Marquardt
\cite{marquardt63}) on MC data to determine the parameters of a Taylor
series expansion of the photon energy in the variables SIZE, WIDTH,
zenith angle and Exp(DIST$^2$). Note
that the latter term takes empirically into account both, shower image
leakage outside the FOV, as well as the drop in light intensity for
impact parameters larger than $100$ m. For the energy reconstruction
studies we used a slightly harder distance cut that does nearly not
affect the collection area below 5 TeV but reduces the collection area
for higher energies by about 25 \%. The
results obtained by this method on a complementary MC data sample are shown in
Figures ~\ref{fig-eresct1}a and ~\ref{fig-eresct1}b.\\
Fig.~\ref{fig-eresct1}a shows the distribution of the relative difference
between initial and the reconstructed energy for a power law spectrum
(differential coefficient: -2.2) above 3.0 TeV. 
Fig.~\ref{fig-eresct1}b shows the predicted energy resolution as function of energy. The worse
energy resolution of CT1 compared to that of CT2 has its origin, besides
the smaller mirror area, in the smaller CT1 camera field of view. At higher energies
a considerable amount of Cherenkov light is falling outside the camera and thus 
information is lost. This loss also affects the angular resolution.
It should be noted that the shown RMS values are 10-30\% larger compared to
standard deviations derived from a gaussian fit.
Fig. ~\ref{fig-eresct1}b also shows the relative deviation between mean
reconstructed energy and initial energy. The deviation is less than 9\% of
the initial energy for both telescopes, but has no effect on the
derived spectra as the unfolding method (see section
\ref{sec-ct1spec}) takes these systematics, both the fluctuation and
the small offset in the energy reconstruction, properly into account.
A detailed description of this energy reconstruction method will be presented
in a forthcoming paper.

\section{Observations and data analysis - CT1}
\label{analysis}
Between March and October 1997 we observed Mkn 501 (ON-source data) with
CT1 for 380 hours at zenith angles between 11$^\circ$ and 60$^\circ$. 
Background data were recorded for a total of 140 h. In order to maximize
ON-source observation time, particularly at small zenith angles, the OFF-source
data were not taken in ON/OFF cycles but
mostly a few hours before or after the Mkn 501 observations. 
Thus not always the equivalent time for
a certain zenith angle setting could be obtained. To compensate for this
deficiency we blended the background at a specific zenith angle range from
data taken at larger and smaller values.
For details (also for the general cutting proceedure) we refer to Petry
(\cite{petry97b}), Kranich (\cite{kranich97b}) and 
Petry \& Kranich (\cite{petrykranich97}).
It should be noted that the observation time was planned well in advance
and that shift operators had no feedback on nightly results such that a
bias to prolonged observation in case of large excess was avoided.

The data analysis proceeded in the following order.
In a first step of data selection the following criteria were applied:
\begin{itemize}
\item[a)]
The atmospheric transmission must be high.
Whenever available, the atmospheric extinction measurements from the
nearby Carlsberg Automatic Meridian Circle (CAMC)
were used. For good data, we require the extinction coefficient in the 
Johnson V-band to be smaller than
$0.25$. 
\item[b)]
The trigger rate, based on 20 min runs, must be within $\pm 10\%$ of the
expected one. This rate is zenith angle dependent.
\item[c)]
Only data up to 38$^\circ$ zenith angle were used for further analysis. 
Due to a lack of MC
events at large zenith angles the data for $\theta_z > 38^\circ$ 
will be analyzed later and presented elsewhere.
\end{itemize}
Since the weather was exceptionally good in 1997 only a few 
nights were lost due to
dense cloud coverage while for the remaining nights always a
high atmospheric transmission was given, see Table 4 for the
Johnson V values (whenever available).
Only 27 hours of data were rejected due to large deviations from the 
expected trigger
rate. Data from 58
hours of observations were deferred for later analysis because the
zenith angle exceeded $38^\circ$.

\begin{figure}
\epsfxsize=8cm
\epsffile{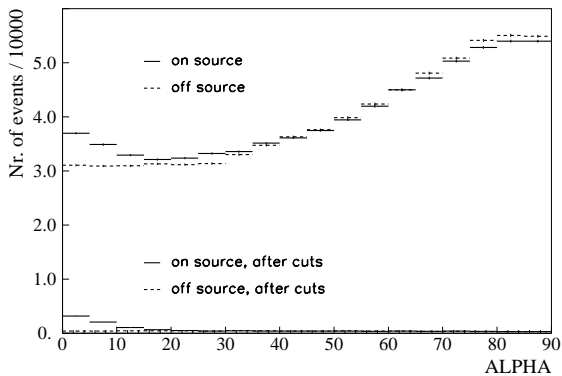}
\caption{\label{fig3}
ALPHA distribution of the raw data after the filter cuts for HV2,
zenith angle $< 38^\circ$ (upper data points)
and the
additional dynamical supercuts (lower data points), see also Fig.~4
for expanded scale.}
\end{figure}

Next, so-called filter cuts were applied rejecting
mostly noise induced triggers. After the FILTER cut the surviving data present a 
nearly pure sample of 
hadronic and $\gamma$ shower images. 
For these events the usual Hillas image parameters were calculated.
Fig.~\ref{fig3} (upper data points)
 shows the ALPHA
distribution for the ON-source data 
as well as for the OFF-source data normalized to the
ALPHA range between 20$^\circ$ and 80$^\circ$. 
Already a clear ON-source excess at
small ALPHA values is seen in the raw data.

After the filter cuts, the data are further reduced by applying the 
above-mentioned dynamical supercuts. These cuts vary with the zenith angle, the
image parameter SIZE (a coarse measure of the initial energy) and
the image parameter DIST (a coarse measure of the impact
parameter). The dynamical cuts enhance significantly the $\gamma$/hadron
($\gamma$/h) ratio. Hadrons are suppressed by a factor 50-60
 while about 60\% of the $\gamma$
showers are retained.

\begin{table}
\centering
\begin{tabular}{lllll}
\hline
HV settings&HV1&HV2&HV3&HV4\\
\hline
Observation time (h)& 104 & 180 & 24.5 & 13.2 \\
Rate after Filter (Hz)& 0.61 & 1.05 & 0.34 & 0.18\\
Excess events & 1875 & 4587 & 206 & 42 \\
Significance ($\sigma$) & 26.2 & 51.1 & 8.7 & 3.9 \\
\hline
\end{tabular}
\caption{Observation times and rates after FILTER for the different HV
  settings for CT1. Data integrated up to 38$^\circ$ zenith angle; but not
  separated for moon/no moon samples.}
\label{ecltab2}
\end{table}

\begin{figure}
\begin{minipage}[b]{8cm}
\epsfxsize=8cm
\epsffile{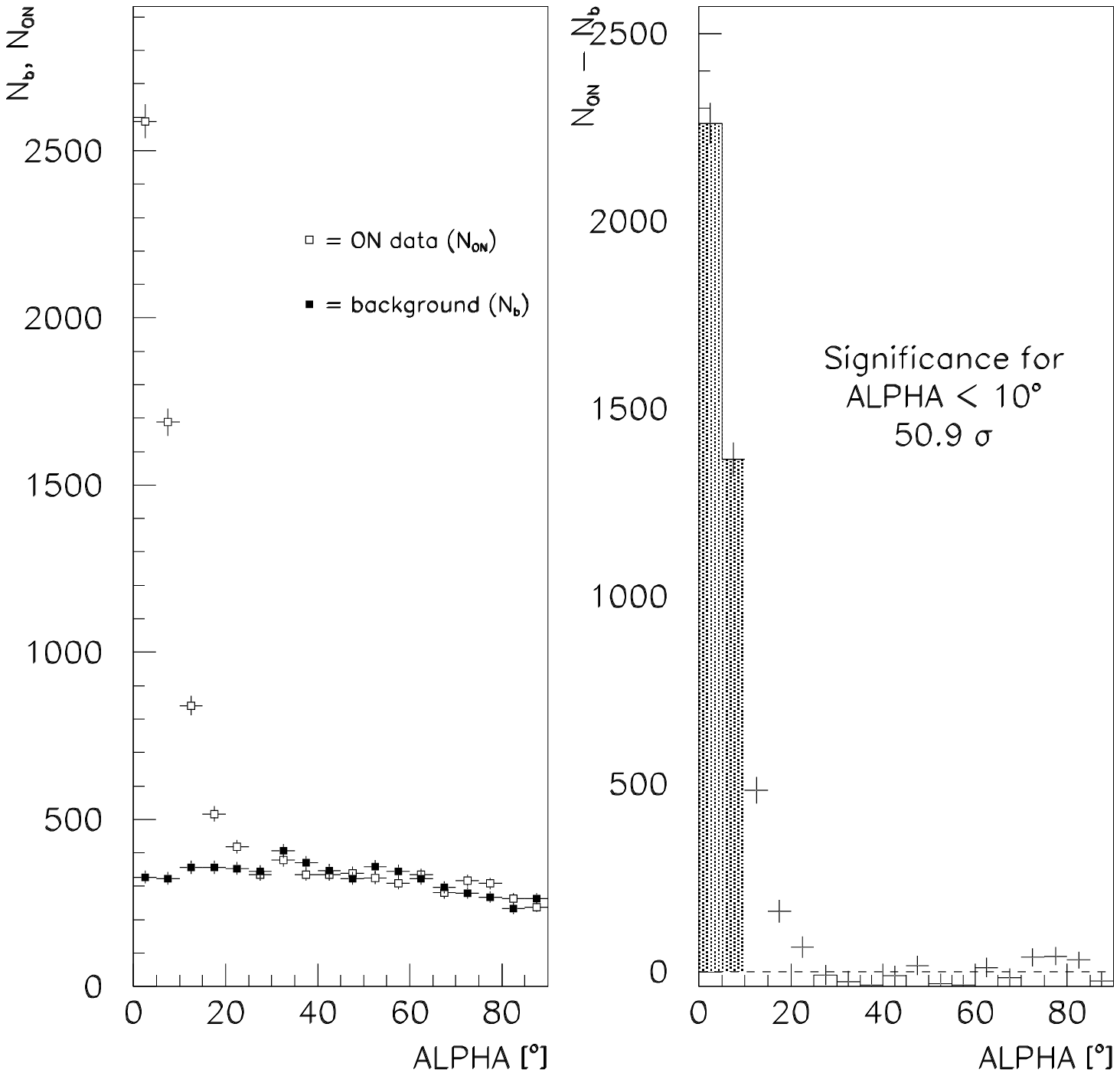}
\end{minipage}
\begin{minipage}[b]{8cm}
\epsfxsize=8cm
\epsffile{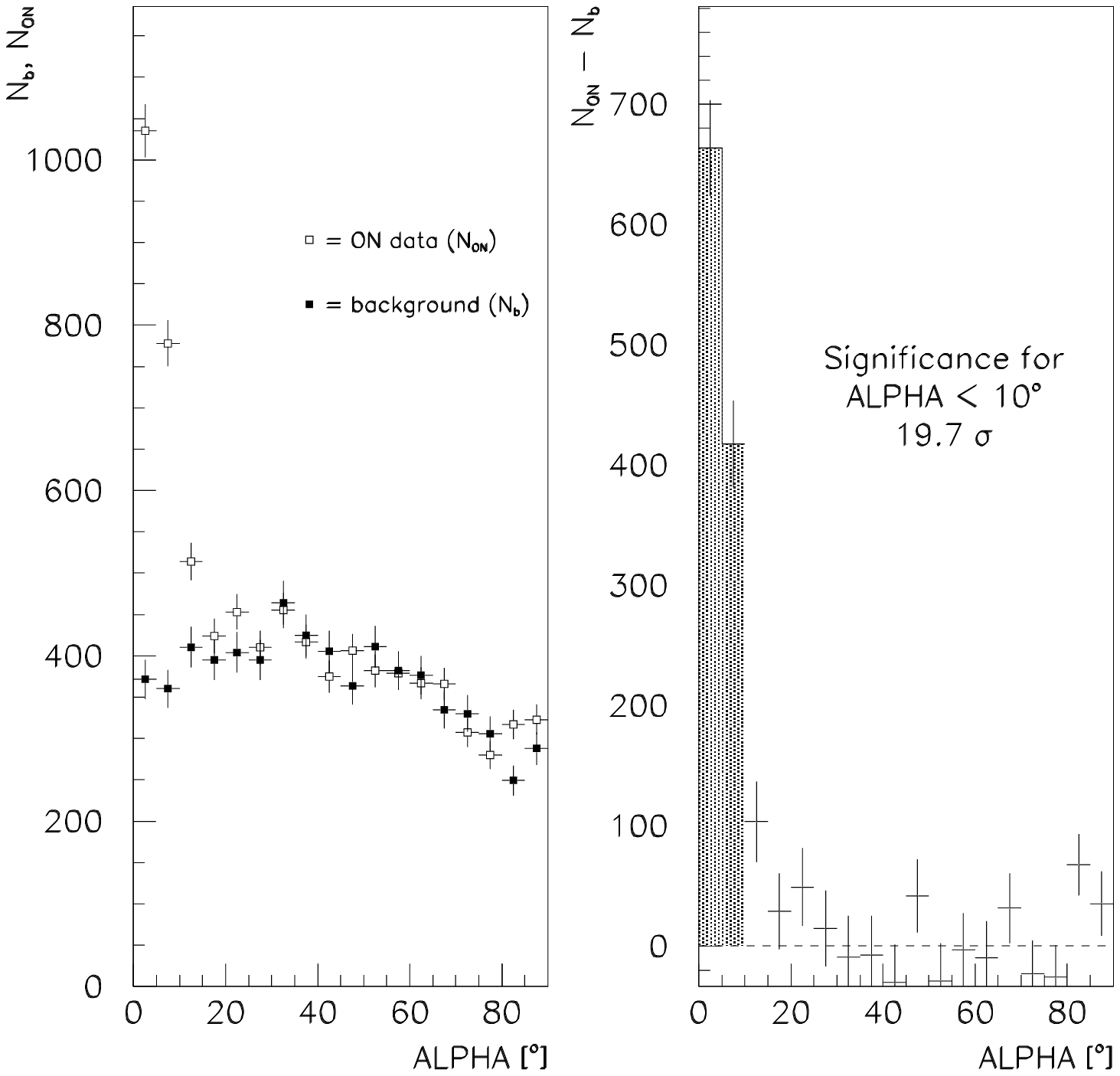}
\end{minipage}
\caption{\label{fig-mkn501ct1}
(a) The total signal from Mkn 501 using all data taken during 
dark nights with HV2 and up to 38$^\circ$ zenith angle;
(b) corresponding distributions with moonshine data
for HV1 (moon $\approx$ 25 - 60\% illuminated).}
\end{figure}

Fig.~\ref{fig-mkn501ct1}a shows the ALPHA distribution for the ON/OFF data
after the dynamical supercuts (HV2, dark nights only). 
The data correspond to 153 hours
ON-source time. Fig.~~\ref{fig-mkn501ct1}b shows the 
equivalent moonlight data for
one HV setting, HV1 (after April 29th).
Table \ref{ecltab2} summarizes for the different HV settings the
observation times and rates for the ON-source data collected with CT1, as
well as the excess signals and significances.

\subsection{Average spectrum}
\label{sec-ct1spec}
In order to derive the energy spectrum from the CT1 data we have used a
technique which implicitly takes into account the effects of the finite energy
resolution. 
 This technique is well known in high energy physics by the name of ``regularised unfolding''
and was developed by Blobel (\cite{blobel84}). 
In brief, this procedure
avoids the oscillating behavior of the solution to unfolding problems by attenuating
insignificant components of the measurements.

The software package ``RUN'' (Blobel \cite{blobel96}) takes three sets of data:
Monte Carlo data, and background data and on-source data after cuts.
From this, it produces - using the regularised unfolding technique - the
corrected fluxes in bins of the energy with a statistical error estimation. 
These values are converted into
differential flux values by dividing by the energy bin width, or into integral
flux values by summing up all contributions above a certain bin number.

Finally, parameters of the spectra are determined by fitting
appropriate functions (see below) to the resulting differential or integral spectrum.

\begin{figure}
\epsfxsize=8cm
\epsffile{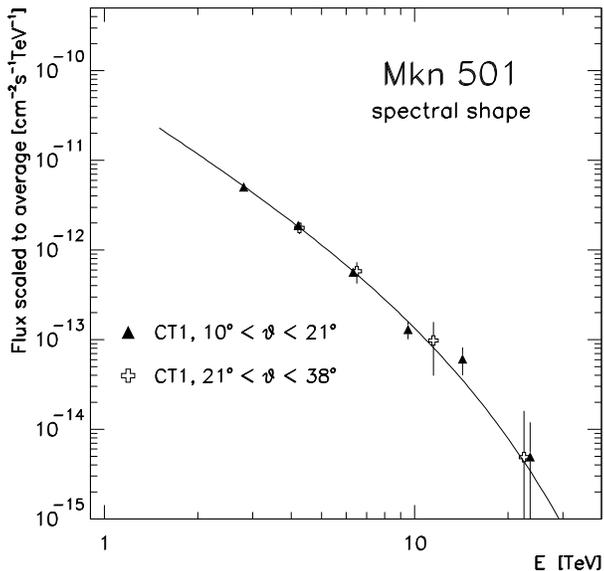}
\caption{\label{fig-ct1spec}
    The average spectral shape as measured by CT1 using the method of regularised unfolding.
    The solid line represents the result from a power law fit with
    exponential cutoff. The errors are purely statistical.}
\end{figure}

In the examination of the spectrum we used only the data from 
dark night observations\footnote{Here we exclude the moonshine-data because
of their higher thresholds making the plot less clear.}.
For the energy estimation for each HV setting, a separate Monte Carlo
simulation was undertaken and an energy reconstruction function
derived (see section \ref{sec_perf_of_ct}).
The energy resolution achieved by this procedure is shown in Figure \ref{fig-eresct1}b.
 After this reconstruction, the data were combined and subdivided 
into two separate zenith angle bins according to the zenith angles of the available
Monte Carlo Data (0$^\circ$ and 30$^\circ$). The first bin  (0$^\circ$-21$^\circ$)
corresponded to 73.0 h observation time, the second (21$^\circ$-38$^\circ$) to
80.2 h observation time. The lowest energy bin for the combined 0$^\circ$-21$^\circ$
data set has its threshold at 2.25 TeV, that for 21$^\circ$-38$^\circ$ at 3.5 TeV.  
The resulting two spectra were scaled such that the fluxes at the point of the lowest
common energy were equal. This was done in order to compensate for the time
variability of the Mkn 501 emission. The result of the unfolding can be seen in Figure 
\ref{fig-ct1spec}. The comparison with the spectral shape from CT2 and
the HEGRA CT system (Part I) is discussed in section \ref{discussion} 
(see also Fig. \ref{all_spec}).

A power law fit to the 10 data points from CT1 yields a differential spectral index of
$$
        \alpha = 2.8 \pm 0.07
$$
with a reduced $\chi^2$ of 1.1. In the concurrently taken data with
the CT system (Part I) and CT2 a significant curvature of the spectrum 
was seen. These data include measurements at much lower energies and
are inconsistent with an unbroken power law. On the other hand the CT1
data are also consistent with the curved spectrum derived from the
system and CT2 data, see discussion in section\ref{discussion}.

\begin{figure}
\epsfxsize=8cm
\epsffile{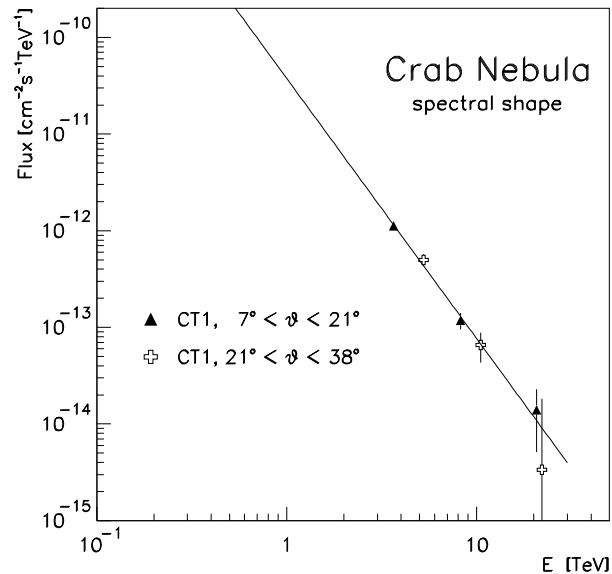}
\caption{\label{fig-crabspec}
    The spectral shape of the Crab Nebula as measured by CT1
    in 1995, 1996 and 1997 (29 h observation time). 
    using the method of regularised unfolding. 
    The superimposed line represents a power law fit.
    All errors are purely statistical.}
\end{figure}

The unfolding method was tested using data on the Crab Nebula. This
data was taken in the years 1995-1997 and amounts to 29 h of
observation time. For comparison the results of this are shown 
in Figure \ref{fig-crabspec}.
A power law fit gives a differential spectral index of 
$$
        \alpha_{\mathrm{Crab}} = 2.69 \pm 0.15
$$
with a $\chi^2$ of 0.5. 
This is in good agreement with other measurements of this source
(Carter-Lewis et al. \cite{carterlewis97}, Konopelko et
al. \cite{konopelko96}, Petry et al. \cite{petry96}, Tanimori et al.
\cite{tanimori98})

\subsection{Average flux}
\label{sec_av_flux}
For a rapidly varying source, such as Mkn 501 in 1997, an averaged flux is
not strictly meaningful because measurements sample the light curve
to only 10-20\% and the observed variability in time is often similar to
the size of the gaps between the daily measurements. Integration over a long 
period of more than 200 days
should nevertheless give a fairly reliable value on the mean flux. In
the following we present the average integral flux above 1.5 TeV. Because
of the various  HV settings and the threshold variation with zenith angle the
threshold was sometimes above 1.5 TeV and extrapolation to 1.5 TeV was
necessary. This was performed using a spectral index of 2.8 as determined
in section \ref{sec-ct1spec}. The systematic error on the integral
flux, arising from the sometimes necessary extrapolation to 1.5 TeV is 
small. Using a simple power law spectrum with differential index of 2.5
yields only a 5\% difference in flux compared to the above spectral
parametrisation.

The signal obtained from CT1 observations has a significance of $\approx 58
\sigma$ (see Table \ref{ecltab2} for the different contributions from the four HV
settings). Therefore the statistical error of the average flux is
completely negligible. Averaging over the four data sets we obtain the
following integral flux above 1.5 TeV
$$
F(E > 1.5\ {\mathrm TeV}) = 2.33 \, (\pm 0.04)_{\rm stat.} \times 10^{-11}
 \mathrm{cm}^{-2}\mathrm{s}^{-1}
$$
between March 11th and October 20th.
This value can be compared with the Crab Nebula flux above 1.5 TeV. From
observations with CT1 in the 1995/96 and 1996/97 winter periods 
(the same dataset as used for Figure \ref{fig-crabspec}), a Crab flux of
$$
F_{\mathrm{Crab}}(E > 1.5\ TeV) = 0.82 \, (\pm 0.1)_{\rm stat.} \times 10^{-11} \mathrm{cm}^{-2}\mathrm{s}^{-1}
$$ 
has been determined (Petry \cite{petry97b}), thus the average flux of 
Mkn 501 in 1997 was about 3 times larger than that of the Crab Nebula. 

The error on the flux is dominated by systematics
and reflects only instrument related errors and not those arising from the
sparse time sampling.
A major contribution to
the error comes from the uncertainty for the photon-to-photoelectron
conversion which we estimated 
to be about 15\%, which, in turn, converts to a systematic flux
uncertainty of $\approx 25\%$. We estimate a total systematic flux error of
30\%, common to all flux values.

\subsection{Test for time variability of the spectral shape}
\label{ct1-specvar}
For the study of Mkn 501's spectral variability above 1.5 TeV we restrict
the analysis to the non-moon data taken at HV2 and zenith angles less than
38$^\circ$ because the thresholds of the individual data sets were 
below 1.5 TeV. For observations
lasting longer than 0.5 hours we calculate daily values of $F_{1.5-3}$, the flux between
1.5 and 3 TeV and $F_3$, the flux above 3.0 TeV. 
The hardness ratio
$$
        r_h = \frac{F_{3}}{F_{1.5-3}}
$$
which is available for over 100 nights, can then be inspected 
for variability.

\begin{figure}
\epsfxsize=8cm
\epsffile{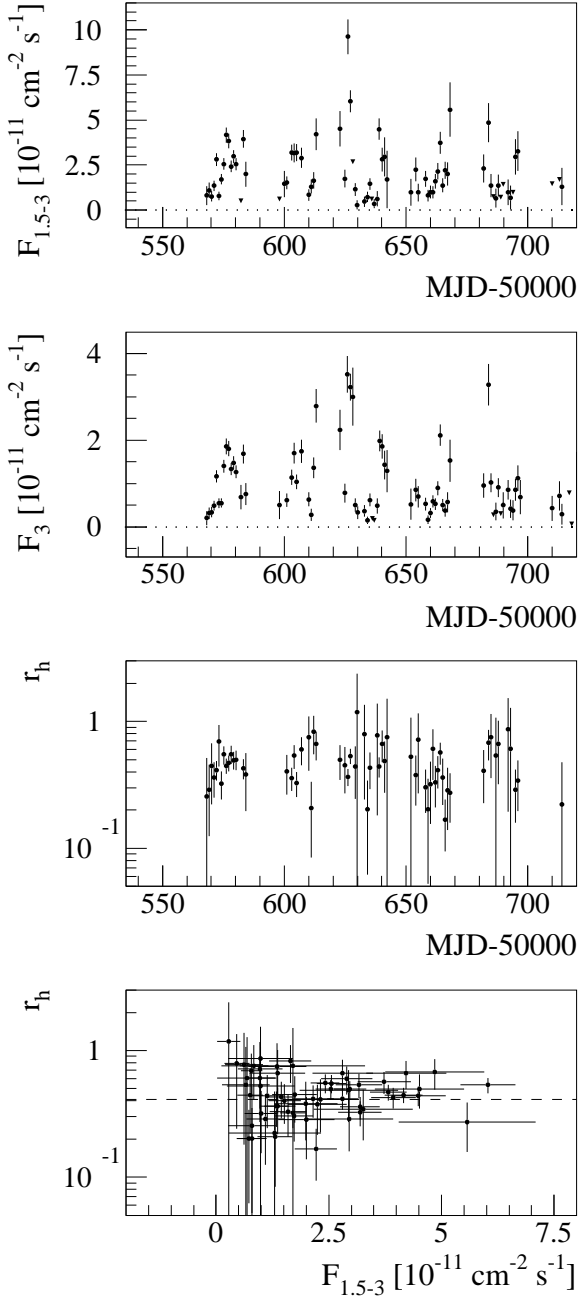}
\caption{\label{fig-ct1vary}
(a) The photon flux $F_{1.5-3}$
of Mkn 501 at energies between 1.5 and 3.0 TeV from the CT1 data taken with HV2.
(b) The photon flux at energies above 3.0 TeV obtained from the 
CT1 data taken with HV2.
(c) The hardness ratio $r_h$ as defined in the text plotted versus time.
(d) The hardness ratio plotted versus the flux between 1.5 and 
3.0 TeV. The fit
  of a constant function gives $r_h = 0.41 \pm 0.02$ and a 
reduced $\chi^2$ of 0.97.}
\end{figure}

Figure \ref{fig-ct1vary} shows the result of this study. 
Only points with significance
$> 1 \sigma$ were used for the calculation of $r_h$ while 
the points with $\le 1 \sigma$
were converted to 90 \% confidence level upper limits and 
are only shown in the light curves.

There is no indication of significant spectral variability 
with time, nor of a correlation
between the hardness ratio and the emission state. The averaged hardness 
ratio of $0.41 \pm 0.02$ (error
purely statistical) is somewhat
smaller than the ratio of $0.51\pm0.01$ as expected
from the spectrum measured by the CT system and CT2 
(section 4),
but the difference is in the
range of the systematic errors. 

In order to estimate the degree of variability still 
permitted by this measurement,
we fit a linear function to the plot $r_h$ versus $F_{1.5-3}$ and obtain
$$
   r_h = (0.006 \pm 0.007 ) \cdot F_{1.5-3} [10^{-11}\mathrm{cm}^{-2}\mathrm{s}^{-1}]
                                     + (0.39 \pm 0.028)
$$
with a reduced $\chi^2$ of 0.98.
With the range of values of $F_{1.5-3}$ of roughly 
$(0.5\, {\rm to}\, 10) \times 10^{-11}\mathrm{cm}^{-2}\mathrm{s}^{-1}$, 
this means that $r_h$ may vary by up to
15\% of its average value within the $1 \sigma$ error of the fit.

\subsection{The CT1 lightcurve above 1.5 TeV}

The lightcurve from CT1 data which we present in this paper 
aims for a time coverage 
as complete as possible and at the same time minimizing 
systematic errors from varying zenith 
angle distributions. In order to achieve this compromise 
we limit the data set to zenith angles of
below 38$^\circ$.
The complete lightcurve of the integral fluxes above 1.5 TeV is shown
in Fig.~\ref{fig-lightcurve}
together with the results from the other HEGRA telescopes.

The lightcurve as shown in 
Figure \ref{fig-lightcurve} is calculated for a threshold of
1.5 TeV. The data are taken from the different HV settings
using the above mentioned extrapolation procedure. 
Only statistical errors are shown. 

The small possible error of $\rm r_h$ up to 15\%, see previous section,
could only influence those points where one has to extrapolate over a
sizable energy range, i.e. for the few data points taken at HV4.
A listing of the CT1 and CT2 observation times and fluxes is given
in Table 4 together with the Johnson V extinction coefficients (whenever
available). Note that the MC simulation takes a mean loss of light of
16\% into account.

\subsection{Correlation with RXTE observations of Mkn 501}
\label{sec-rxtecorr}
Since the beginning of 1996, the RXTE all sky monitor (ASM) 
has been observing Mkn 501
in the 2-12 keV band. From these data (subsequently called {\it keV} data),
which are publicly available as so-called ``quick-look results'', the
hardness ratio

\begin{displaymath}
\frac{\mathrm{Rate} (5-12.1\ \mathrm{keV})}
{\mathrm{Rate} (1.3-3\ \mathrm{keV})}\,
\end{displaymath}
has been determined.

\begin{figure}
\epsfxsize=8cm
\epsffile{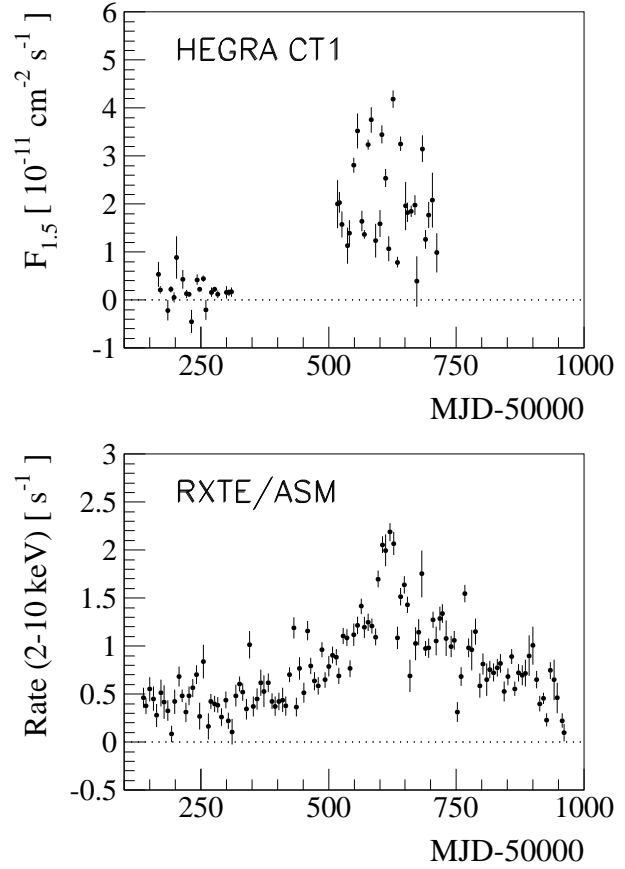}
\caption{\label{fig-plot2years}
        Observations of Mkn 501 in 1996 and 1997 at energies above 1.5 TeV (using HEGRA CT1)
        compared to the X-ray emission between 2 and 10 keV as measured by RXTE ASM. In both
        plots, each point represents the emission averaged over a period of 7 days.}
\end{figure}

In Figure \ref{fig-plot2years}, we present the RXTE keV rate together with the data from HEGRA CT1 which observed Mkn 501
both in 1996 and 1997. The simultaneous change in flux in both energy
ranges is clearly visible.

\begin{figure}
\epsfxsize=8cm
\epsffile{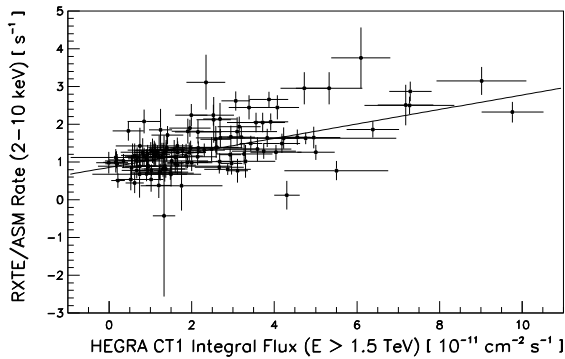}
\caption{\label{fig-rxtecorr_noshift}
        Correlation between the X-ray emission of Mkn 501 in 1997 as measured by RXTE
        and the emission above 1.5 TeV as measured by CT1. A linear fit to the data points yields:
        RXTE Rate $= (0.20 \pm 0.01) \times \mathrm{CT1} + (0.86 \pm 0.04)$\,Hz, 
        where CT1 represents the flux above 1.5 TeV in units of $10^{-11}$\,cm$^{-2}$s$^{-1}$.}
\end{figure}

In order to further examine the correlation, we plot the daily RXTE averages versus
the flux values from the complete CT1 lightcurve in 1997. This is 
shown in Fig.
\ref{fig-rxtecorr_noshift}. We obtain a correlation coefficient 
(see Part I for details) of:
$$
        r = 0.611 \pm 0.057.
$$
with a significance of $8.56$ (based on the assumption of 125 independent data
pairs).

\begin{figure}
\epsfxsize=8cm
\epsffile{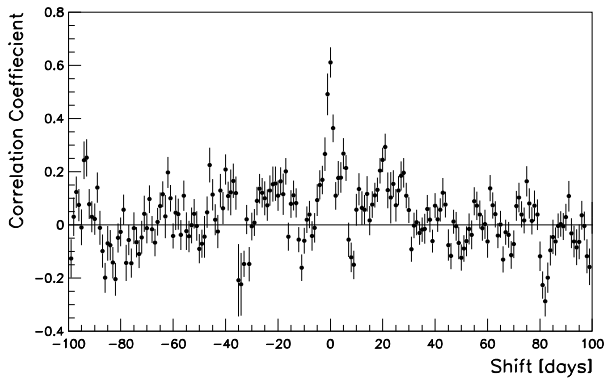}
\caption{\label{fig-rxtecorr_shift}
        Check of the correlation between the X-ray emission of Mkn 501 in 1997 as measured by RXTE
        and the emission above 1.5 TeV as measured by CT1. Plotted is the correlation coefficient
        as a function of the artificially introduced shift between the RXTE and the CT1 time.}
\end{figure}

In order to verify whether this correlation is real or only an
artifact of some binning effects (e.g. fewer observations during
moonshine nights, etc.) or unequal data statistics, we shift the CT1 and the RXTE light curves
with respect to each other in steps of 1 day by up to $\pm 100$
days. For each shift, we recalculate the correlation coefficient. The
result is plotted in Figure \ref{fig-rxtecorr_shift}. The fact that
only at the un-shifted value there is a clear peak visible, underlines
the significant correlation between the TeV and keV datasets. Even if we assume 
that the daily TeV data are highly correlated, i.e., only $1\, /\, 5$ of the data
is independent, we obtain still a significance of $3.7$.
 

Due to the nearly continuous CT1 observation only a modest
 modulation due to
the lunar period is visible. 

The ratio
\begin{displaymath}
R_{\rm TeV/keV} = \frac{\langle F_{1.5}\rangle}{\langle {\rm
    RXTE~countrate}\rangle} 
\end{displaymath}
(here in units of [$10^{-11}$ cm$^{-2}$s$^{-1}$/Hz]) is quite different
in 1996 and 1997: 
\begin{displaymath}
R_{\rm TeV/keV} ({\rm MJD}\ 50160 - 50310) = 0.5 \pm 0.04
\end{displaymath}
and
\begin{displaymath}
R_{\rm TeV/keV} ({\rm MJD}\ 50520 - 50720) = 1.75 \pm 0.04.
\end{displaymath}
While the keV flux rises by about
a factor 3 from the 1996 to 1997 periods, 
the TeV $\gamma$ flux increases 
by 11, i.e.,
about in quadrature of the keV flux.

\section{Observations and data analysis - CT2}
CT2 observed Mkn 501 between 16 March and 28 August 1997. After thorough checks
of the data quality, 85 hours ( 79 {\it h} in normal- and 6 {\it h} in reverse
tracking mode \footnote{Normal- and reverse mode refer to the Azimuth
range the Telescope is operated when observing a source. The Azimuth
range of the reverse mode refers to a $180^\circ$ rotation in
$\varphi$} rotation of the telescope) of good data remained. 

For the background determination the same procedure was followed as for CT1..  
The OFF-source data set consisted
of 90 hours of data which had passed the same quality cuts as the ON-data and
spanned all zenith angles up to 51$^\circ$.

For the gamma/hadron separation,  we employed the
set of image parameter cuts already used in 
Petry (\cite{petry96}). The efficiencies of these
cuts and the corresponding Monte Carlo studies are described in 
Bradbury et al. (\cite{bradbury97}) and Petry (\cite{petry97b}).
The effective collection areas of CT2 after gamma/hadron separation cuts
 for three different zenith angles are shown 
in Figure \ref{fig-CT1-CT2}d. 
The characteristics of CT2 have not changed over a long time. This was checked by comparing
data from Mkn 421 observations taken in 1995 with the 1997 Mkn 501 data set. Neither
the background rates nor the background image parameter distributions of CT2 have 
changed significantly. 

 Table \ref{tab-ct2} summarizes the observation times and
 trigger rates before and after the FILTER cut for CT2 for 3 ranges of the
 zenith angle. Also given are the excess and background rates and signal
 significances after the ``image'' cuts.
Due to the coarser camera pixel size, an ALPHA cut at 
15$^\circ$ is applied.

\begin{figure}
\centering
\epsfxsize=8cm
\epsffile{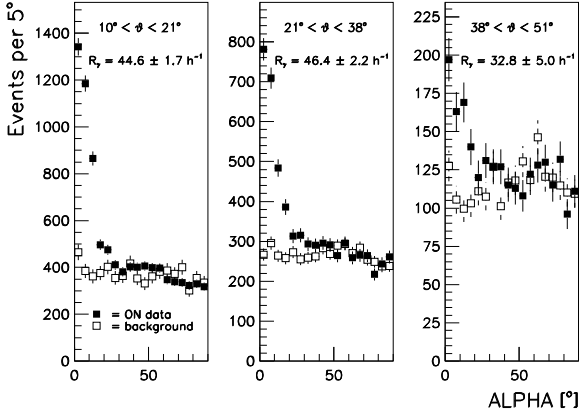}
\caption{\label{fig-dreiza_alphaplot}
        The ALPHA distributions after all cuts for all normal-tracking data from Mkn 501 observations
        with CT2 in 1997, split up in three zenith angle ($\vartheta$) bins. $R_{\gamma}$
        denotes time averaged rates.}
\end{figure}

Fig.~\ref{fig-dreiza_alphaplot} shows the CT2 ALPHA distributions after
all cuts, for the zenith angle ranges as listed in Table \ref{tab-ct2}. In
all distributions a clear excess at small ALPHA is seen.

\begin{table*}
\begin{center}
\begin{tabular}{llcccc}
                
                zenith angle range & $0^\circ - 21 ^\circ$ & $21^\circ - 38^\circ$ & $38^\circ - 51^\circ$ & total \\
                \hline
Observation time (h)& 48.3 & 24.7 & 6.0 & 79\\                       
  Av. Rate before filter (Hz) &  2.57 &  2.62 & 2.54 & \\      
 Av. Rate after filter (Hz)$^{a)}$ &  1.40 & 1.35 & 0.85 & \\
background (ALPHA $< 15^\circ$)&1211&862.2&332.5&\\
excess signal (ALPHA $< 15^\circ$)& 2181&1147&196.5&\\
significance$^{b)}$&25.7$\sigma$&21.06 $\sigma$&6.6$\sigma$&34\\
                \hline
        \end{tabular}
\end{center}
        \caption{\label{tab-ct2}
The CT2 data sets taken in normal-tracking mode
(as used in Fig. \ref{fig-dreiza_alphaplot},
\ref{fig-ct2spec} and \ref{fig-ct2vary}) for 3 zenith angle ranges.
a) The large reduction by the filter cuts has its origin in fake
  triggers from muons passing the Plexiglas focons of the PM camera and in the 
  exclusion of the outer pixel ring from the software trigger.
b) The quoted significance is slightly different from a value calculated directly
from the given excess and background rate. The difference has its origin in
a somewhat different background density population as a function of zenith
angle.}
\end{table*}

\subsection{Average spectrum and flux}
\label{CT2spec}

For the study of the spectrum of the CT2 signal we applied the
regularised unfolding method as for CT1 (see section \ref{sec-ct1spec}).
We  subdivided the dataset into 
three separate zenith angle bins according to the zenith angles of the available
Monte Carlo data (0$^\circ$, 30$^\circ$ and 45$^\circ$)
(see Table \ref{tab-ct2} for statistics).


\begin{figure}
\epsfxsize=8cm
\epsffile{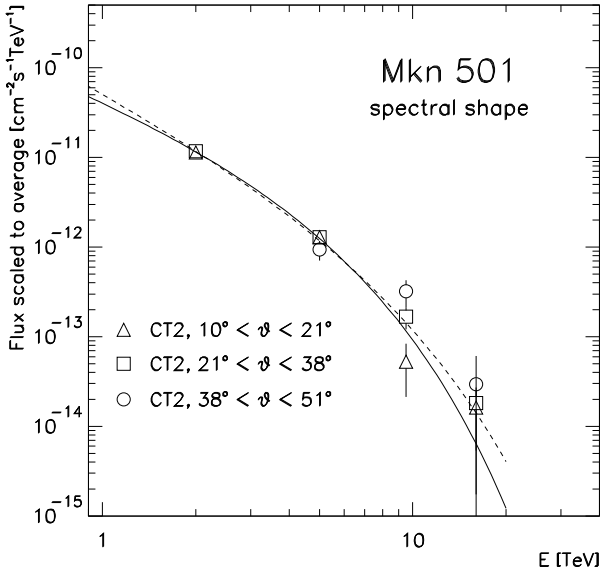}
\caption{\label{fig-ct2spec}
    The average spectral shape as measured by CT2 using the method of
    regularised unfolding. The curves represent the results
    from the following fits: Solid line: fit of a power law with
    exponential cutoff, Dashed line: fit of a power
    law with fixed index 1.8 and exponential cutoff.}
\end{figure}

For each of these datasets the regularised unfolding was applied separately.
The resulting three spectra were scaled such that the fluxes at the point of the lowest 
common energy were equal. This was done in order to compensate for the
flux variation with time.
The results are shown in Figure \ref{fig-ct2spec}.Within the errors,
the spectra from different zenith angle observations are perfectly compatible.

When fitting the CT2 data by 
a pure power law, we obtain a differential spectral index of
$$
        \alpha = 2.7 \pm 0.2
$$
with a reduced $\chi^2$ of 5.0. In order to improve the fit, we introduce an 
exponential cutoff, i.e. we fit
$$
        {\mathrm d}F/{\mathrm d}E \propto E^{-\alpha} \cdot e^{-E/E_0}.
$$
This fit gives
$$
        \alpha = 1.27 \pm 0.37, E_0 = (2.85 \pm 0.58) \mathrm{TeV}
$$      
with a reduced  $\chi^2$ of 1.7. This is shown in Figure \ref{fig-ct2spec}
as a solid line.

Using the CT2 data and the above spectrum (when an extrapolation was
necessary), we calculate an average flux above 1 TeV 
\begin{displaymath}
F (E > 1.0\ {\rm TeV}) = 5.26 \, (\pm 0.13)_{\rm stat.}
 \times 10^{-11}{\rm cm}^{-2}{\rm sec}^{-1}
\end{displaymath}
for the 85 hours of observation time.

The corresponding Crab flux value measured with CT2 is (Petry et al. 
\cite{petry96}):
\begin{eqnarray*}
F^{\rm Crab} (E > 1.0\ {\rm TeV}) &=& 1.57 \, (\pm 0.24)_{\rm stat.}
(+ 0.99 - 0.39)_{\rm syst.} \\
  && \times 10^{-11}{\rm cm}^{-2}{\rm sec}^{-1}\, .
\end{eqnarray*}
Here we use the Crab flux as determined by CT2 and not with CT1 because the
ratio $F^{\rm Mkn~501}/F^{\rm Crab}$, when
 measured with the same telescope,
 should be free of some systematic
errors, such as the photon to ADC signal conversion error.
The resulting flux ratio $F^{\rm Mkn~501}/F^{\rm Crab}$ is $3.3\pm
0.5$ and in good agreement with the ratio obtained from the CT1 data
(sec. \ref{sec_av_flux}).

\subsection{Test for time variability of the spectral shape}
As for CT1 (section \ref{ct1-specvar}), we examine the possible spectral variability
of Mkn 501 in the independent CT2 data set. Again, we
 construct from Monte Carlo data a function 
which estimates the energy of the primary photon from the zenith angle and the
image parameters SIZE, DIST and WIDTH (see also Fig. \ref{fig-eresct1}b).

For the daily measurements lasting longer than 0.5 hours, 
we determine the flux $F_{1-3}$ between 
1.0 and 3.0 TeV and the
flux $F_3$ above 3.0 TeV and define a hardness ratio $r_h$ as
$$
        r_h = \frac{F_3}{F_{1-3}}.
$$
Note that this is different from the hardness ratio defined for CT1 since the threshold of
CT2 is lower. Only data up to a zenith angle of 30$^\circ$ were used.

\begin{figure}
\epsfxsize=8cm
\epsffile{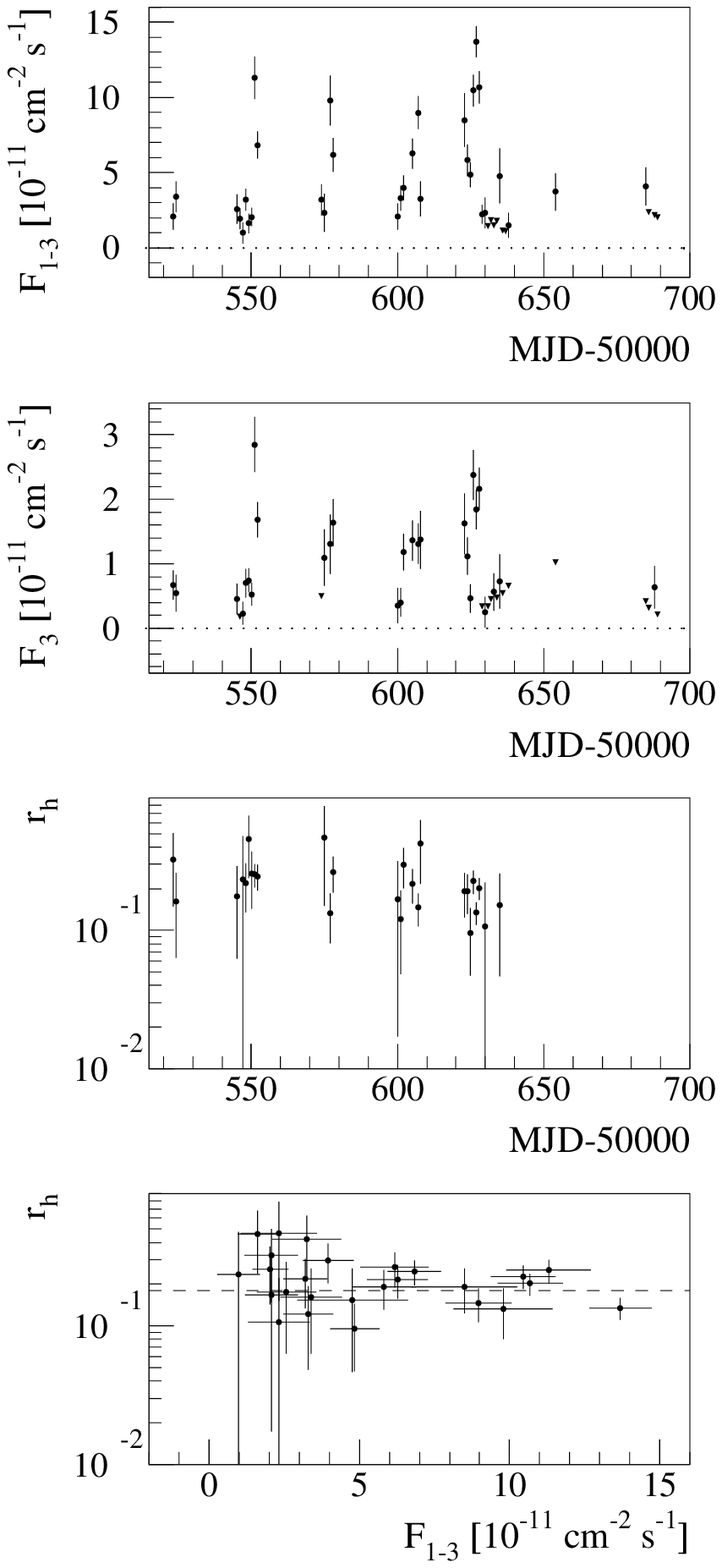}
\caption{\label{fig-ct2vary}
(a) The photon flux of Mkn 501 at energies between 1.0 and 3.0 TeV determined from the CT2 data.
(b) The photon flux at energies above 3.0 TeV obtained from the CT2 data.
(c) The hardness ratio $r_h$, as defined in the text, plotted versus time.
(d) The hardness ratio plotted versus the flux between 1.0 and 3.0 TeV. The fit
  of a constant function gives $r_h = 0.18\pm0.012$ and a reduced $\chi^2$ of 0.92.}
\end{figure}

In Figure \ref{fig-ct2vary} we plot $F_{1-3}$, $F_3$ and $r_h$ versus time and in addition
$r_h$ versus $F_{1-3}$ to test a dependence on the emission state of the source.
A fit of a constant function to the latter plot results in
$$
        r_h = 0.18 \pm 0.012
$$
(errors purely statistical) 
which is in agreement with the value $0.24 \pm 0.02$ expected from the measured 
spectrum if we take into account the large systematic errors of the energy calibration
of $\approx$ 20 \%. given the good reduced $\chi^2$ of 0.92, there
is no indication for a correlation between the hardness ratio and the
emission state.

As for the corresponding CT1 data, we give an estimate of the degree of variability 
still permitted by fitting a linear function to the plot $r_h$ versus $F_{1-3}$. We obtain
\begin{eqnarray*}
 r_h &=& (-0.0038 \pm 0.0032 ) \cdot F_{1-3} [10^{-11}\mathrm{cm}^{-2}\mathrm{s}^{-1}]\\
                      &&               + 0.215 \pm 0.032.
\end{eqnarray*}
(Note again that $r_h$ is differently defined for CT1 and CT2).
With the range of values of $F_{1-3}$ of roughly 
$(1-14) \times 10^{-11}\mathrm{cm}^{-2}\mathrm{s}^{-1}$, this means that $r_h$ may vary by up to
25 \% of its average value within the $1 \sigma$ error bars of the fit. However, the fact that the
slope in the corresponding result of CT1 has the opposite sign, is an indication that
indeed no spectral variability is present.

\subsection{The CT2 lightcurve above 1.5 TeV}
\label{CT2light}
In order to examine the time variability of the emission of Mkn 501
and to compare the data with those of the other telescopes, we construct
the lightcurve above 1.5 TeV. We exclude again the data with
zenith angles larger than 38$^\circ$ in order to avoid possible systematic errors due to
low MC statistics at large zenith angles.

\begin{figure*}
\epsfxsize=17cm
\epsfysize=21cm
\epsffile{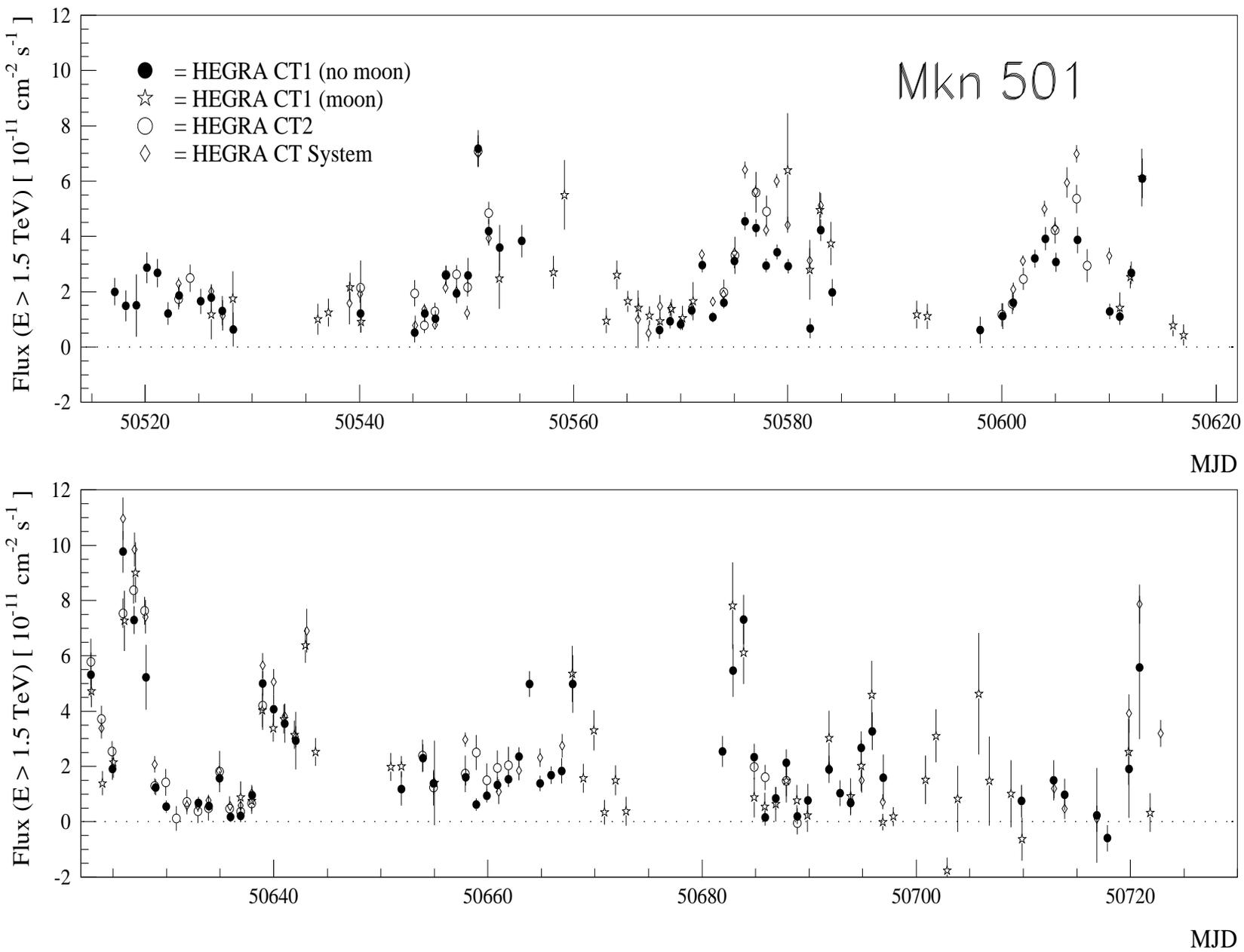}
\caption{\label{fig-lightcurve}
  The measurements of CT1 at all HV settings, with and without the presence of moonlight,
  and of CT2. All points are scaled to the threshold of 1.5 TeV using the spectrum measured
  in section \protect\ref{sec-ct1spec} and \protect\ref{CT2spec}. The measurements of the CT system (Part I)
are added for comparison.}
\end{figure*}

Each point is calculated according to the method of the 
adjustment of the zenith angle distribution (see section \ref{analysis}).
The errors  are purely statistical. 
The results are shown in Fig.~\ref{fig-lightcurve} as open circles.

CT2 observations were partly overlapping in time
with CT1 or the CT system and partly carried out
alone. Therefore the CT2 measurements provide cross checks of the CT1
and CT system measurements and they add some new data points to the
lightcurve. The daily fluxes and
observation times are listed in Table \ref{CT2 flux data} , again for zenith angles below
38$^\circ$ and $E > 1.5$ TeV.

\begin{table}
\begin{tabular}{|c|c|c|c|}
\hline
 Start & $ \Phi \left( \, {\rm E} > 1.5 \, {\rm TeV} \, \right) $ & Duration & Extinction \\
  $[\, \rm MJD \,]$ & $[\, 10^{-11} {\rm cm^{-2} s^{-1}}]\,$  & $[\, \rm h \,]$ & $[\,\rm Johnson V \,]$\\
\hline
   50517.16   & $   2.00  \pm  0.50 $  &   1.82 &   \\
   50518.17   & $   1.49  \pm  0.57 $  &   1.35 &   \\
   50519.16   & $   1.50  \pm  1.13 $  &   0.45 &   \\
   50520.16   & $   2.87  \pm  0.57 $  &   1.87 &   \\
   50521.15   & $   2.67  \pm  0.52 $  &   2.11 &   \\
   50522.15   & $   1.21  \pm  0.41 $  &   2.08 &   \\
   50523.15   & $   1.86  \pm  0.41 $  &   2.47 &   \\
   50525.17   & $   1.65  \pm  0.46 $  &   1.65 &   \\
   50526.16   & $   1.66  \pm  0.41 $  &   2.61 &   \\
   50527.21   & $   1.31  \pm  0.53 $  &   0.92 &   \\
   50528.20   & $   0.94  \pm  0.53 $  &   1.15 &   \\
   50536.11   & $   1.00  \pm  0.56 $  &   2.45 & 0.15  \\
   50537.11   & $   1.24  \pm  0.51 $  &   2.97 & 0.14  \\
   50539.12   & $   2.16  \pm  0.53 $  &   2.82 &   \\
   50540.13   & $   0.98  \pm  0.34 $  &   2.87 & 0.13  \\
   50545.18   & $   0.53  \pm  0.37 $  &   1.24 &   \\
   50546.09   & $   1.20  \pm  0.31 $  &   3.41 & 0.10  \\
   50547.08   & $   1.02  \pm  0.30 $  &   3.50 & 0.11  \\
   50548.08   & $   2.58  \pm  0.37 $  &   3.56 &   \\
   50549.07   & $   1.94  \pm  0.35 $  &   3.61 &   \\
   50550.16   & $   2.59  \pm  0.64 $  &   1.20 & 0.12  \\
   50551.07   & $   7.18  \pm  0.68 $  &   2.53 & 0.12  \\
   50552.07   & $   4.18  \pm  0.44 $  &   3.65 & 0.11  \\
   50553.09   & $   3.19  \pm  0.67 $  &   1.58 &   \\
   50555.15   & $   3.83  \pm  0.59 $  &   1.66 &   \\
   50558.13   & $   2.70  \pm  0.60 $  &   2.27 &   \\
   50559.14   & $   5.50  \pm  1.26 $  &   0.97 &   \\
   50563.04   & $   0.96  \pm  0.46 $  &   4.10 & 0.13  \\
   50564.04   & $   2.61  \pm  0.52 $  &   4.01 &   \\
   50565.05   & $   1.65  \pm  0.38 $  &   3.99 & 0.13  \\
   50566.05   & $   1.41  \pm  0.39 $  &   3.57 & 0.15  \\
   50567.09   & $   1.13  \pm  0.35 $  &   2.99 & 0.12  \\
   50568.07   & $   0.78  \pm  0.23 $  &   4.30 & 0.13  \\
   50569.07   & $   1.10  \pm  0.22 $  &   4.29 &   \\
   50570.06   & $   0.86  \pm  0.19 $  &   4.23 & 0.13  \\
   50571.04   & $   1.35  \pm  0.20 $  &   4.37 & 0.11  \\
   50572.02   & $   2.96  \pm  0.26 $  &   4.53 & 0.12  \\
   50573.02   & $   1.08  \pm  0.19 $  &   4.18 & 0.13  \\
   50574.02   & $   1.60  \pm  0.20 $  &   4.69 &   \\
   50575.01   & $   3.11  \pm  0.26 $  &   4.78 & 0.10  \\
   50576.02   & $   4.55  \pm  0.33 $  &   4.27 & 0.13  \\
   50577.02   & $   4.30  \pm  0.32 $  &   4.35 & 0.11  \\
   50578.00   & $   2.95  \pm  0.25 $  &   4.93 & 0.13  \\
   50579.00   & $   3.43  \pm  0.28 $  &   4.53 & 0.13  \\
   50580.01   & $   2.98  \pm  0.26 $  &   4.53 &   \\
   50582.07   & $   0.89  \pm  0.35 $  &   1.19 &   \\
   50583.04   & $   4.42  \pm  0.34 $  &   4.86 & 0.13  \\
   50584.08   & $   2.47  \pm  0.42 $  &   2.27 &   \\
   50592.05   & $   1.17  \pm  0.51 $  &   3.46 & 0.12  \\
   50593.05   & $   1.11  \pm  0.47 $  &   3.49 &   \\
   50597.97   & $   0.62  \pm  0.48 $  &   0.49 &   \\
   50600.05   & $   1.11  \pm  0.47 $  &   0.50 &   \\
   50601.06   & $   1.59  \pm  0.28 $  &   2.50 & 0.11  \\
   50603.05   & $   3.20  \pm  0.34 $  &   2.99 &   \\
   50604.07   & $   3.91  \pm  0.42 $  &   2.41 & 0.16  \\
   50605.04   & $   3.07  \pm  0.35 $  &   3.00 & 0.12  \\
   50607.07   & $   3.87  \pm  0.48 $  &   1.92 & 0.11  \\
\hline
\end{tabular}
\\(table continues)
\end{table}

 \begin{table}
\begin{tabular}{|c|c|c|c|}
\hline
 Start & $ \Phi \left( \, {\rm E} > 1.5 \, {\rm TeV} \, \right) $ & Duration & Extinction \\
  $[\, \rm MJD \,]$ & $[\, 10^{-11} {\rm cm^{-2} s^{-1}}]\,$  & $[\, \rm h \,]$ & $[\,\rm Johnson V \,]$\\
\hline
   50610.05   & $   1.28  \pm  0.28 $  &   2.41 & 0.12  \\
   50611.02   & $   1.16  \pm  0.26 $  &   1.99 &   \\
   50612.04   & $   2.61  \pm  0.29 $  &  3.44   &   \\
   50613.09   & $   6.10  \pm  0.60 $  &  1.89   &   \\
   50615.98   & $   0.78  \pm  0.70 $  &  3.97   & 0.12  \\
   50616.96   & $   0.43  \pm  0.38 $  &  4.32   & 0.12  \\
   50622.95   & $   4.93  \pm  0.47 $  &  3.82   &   \\
   50623.99   & $   1.39  \pm  0.45 $  &  2.31   & 0.11  \\
   50624.98   & $   1.99  \pm  0.32 $  &  2.99   & 0.11  \\
   50625.98   & $   8.96  \pm  0.62 $  &  2.49   & 0.10  \\
   50626.98   & $   7.61  \pm  0.47 $  &  3.46   &   \\
   50628.07   & $   5.23  \pm  1.18 $  &  0.50   & 0.11  \\
   50628.96   & $   1.25  \pm  0.29 $  &  2.00   & 0.11  \\
   50629.95   & $   0.54  \pm  0.23 $  &  1.84   &   \\
   50632.95   & $   0.68  \pm  0.23 $  &  2.00   & 0.14  \\
   50633.95   & $   0.57  \pm  0.19 $  &  3.00   & 0.12  \\
   50634.95   & $   1.58  \pm  0.26 $  &  3.00   & 0.12  \\
   50635.95   & $   0.18  \pm  0.19 $  &  2.50   &   \\
   50636.93   & $   0.27  \pm  0.17 $  &  3.33   & 0.11  \\
   50637.98   & $   0.86  \pm  0.24 $  &  2.32   & 0.20  \\
   50638.95   & $   4.67  \pm  0.37 $  &  3.68   & 0.19  \\
   50639.98   & $   3.71  \pm  0.36 $  &  3.44   & 0.12  \\
   50640.98   & $   3.66  \pm  0.43 $  &  3.04   & 0.16  \\
   50641.95   & $   3.10  \pm  0.47 $  &  2.76   & 0.11  \\
   50642.94   & $   6.38  \pm  0.64 $  &  2.80   & 0.14  \\
   50643.91   & $   2.52  \pm  0.50 $  &  3.05   & 0.13  \\
   50650.94   & $   1.98  \pm  0.51 $  &  3.55   & 0.12  \\
   50651.94   & $   1.76  \pm  0.33 $  &  3.24   & 0.12  \\
   50653.93   & $   2.31  \pm  0.50 $  &  1.00   &   \\
   50654.94   & $   1.39  \pm  0.40 $  &  1.30   &   \\
   50657.92   & $   1.62  \pm  0.31 $  &  2.45   & 0.14  \\
   50658.90   & $   0.62  \pm  0.22 $  &  2.80   & 0.12  \\
   50659.90   & $   0.94  \pm  0.25 $  &  2.79   & 0.11  \\
   50660.90   & $   1.33  \pm  0.27 $  &  2.76   & 0.12  \\
   50661.90   & $   1.54  \pm  0.29 $  &  2.76   &   \\
   50662.90   & $   2.36  \pm  0.34 $  &  2.67   &   \\
   50663.90   & $   4.98  \pm  0.47 $  &  2.60   &   \\
   50664.90   & $   1.39  \pm  0.28 $  &  2.49   & 0.16  \\
   50665.89   & $   1.69  \pm  0.32 $  &  2.00   & 0.18  \\
   50666.91   & $   1.84  \pm  0.46 $  &  1.00   & 0.11  \\
   50667.91   & $   5.17  \pm  0.73 $  &  1.33   &   \\
   50668.89   & $   1.57  \pm  0.52 $  &  1.49   & 0.11  \\
   50669.90   & $   3.30  \pm  0.73 $  &  1.50   & 0.11  \\
   50670.89   & $   0.35  \pm  0.45 $  &  1.50   & 0.11  \\
   50671.89   & $   1.50  \pm  0.54 $  &  2.24   &   \\
   50672.89   & $   0.39  \pm  0.53 $  &  2.19   & 0.11  \\
   50681.89   & $   2.54  \pm  0.57 $  &  1.00   & 0.10  \\
   50682.87   & $   6.11  \pm  0.82 $  &  1.49   & 0.10  \\
   50683.88   & $   6.85  \pm  0.71 $  &  1.50   &   \\
   50684.87   & $   1.89  \pm  0.40 $  &  2.12   &   \\
   50685.88   & $   0.26  \pm  0.27 $  &  1.98   &   \\
   50686.88   & $   0.79  \pm  0.35 $  &  1.33   &   \\
   50687.88   & $   1.97  \pm  0.41 $  &  1.95   & 0.10  \\
   50688.88   & $   0.32  \pm  0.26 $  &  1.87   & 0.10  \\
   50689.87   & $   0.49  \pm  0.43 $  &  0.82   & 0.09  \\
   50691.88   & $   2.13  \pm  0.45 $  &  1.65   & 0.08  \\
   50692.89   & $   1.03  \pm  0.47 $  &  1.00   &   \\
   50693.87   & $   0.75  \pm  0.37 $  &  1.16   & 0.12  \\
\hline
\end{tabular}
\\(table continues)
\end{table}

\begin{table}
\begin{tabular}{|c|c|c|c|}
\hline
 Start & $ \Phi \left( \, {\rm E} > 1.5 \, {\rm TeV} \, \right) $ & Duration & Extinction \\
  $[\, \rm MJD \,]$ & $[\, 10^{-11} {\rm cm^{-2} s^{-1}}]\,$  & $[\, \rm h \,]$ & $[\,\rm Johnson V \,]$\\
\hline
   50694.88   & $   2.48  \pm  0.50 $  &  1.49   &   \\
   50695.87   & $   3.58  \pm  0.60 $  &  1.43   & 0.12  \\
   50696.87   & $   0.17  \pm  0.29 $  &  1.42   & 0.14  \\
   50697.86   & $   0.19  \pm  0.35 $  &   1.35  & 0.25  \\
   50700.87   & $   1.52  \pm  0.88 $  &   1.01  &   \\
   50701.85   & $   3.10  \pm  0.96 $  &   1.23  &   \\
   50702.88   & $  -1.75  \pm  0.46 $  &   0.46  &   \\
   50703.85   & $   0.83  \pm  1.19 $  &   1.13  &   \\
   50705.86   & $   4.63  \pm  2.20 $  &   0.73  &   \\
   50706.84   & $   1.47  \pm  1.62 $  &   0.66  &   \\
   50708.86   & $   1.02  \pm  1.21 $  &   0.79  &   \\
   50709.85   & $   0.26  \pm  0.47 $  &   0.98  &   \\
   50712.85   & $   1.50  \pm  0.74 $  &   0.69  &   \\
   50713.84   & $   0.98  \pm  0.57 $  &   0.78  &   \\
   50716.86   & $   0.24  \pm  1.72 $  &   0.12  &   \\
   50717.85   & $  -0.59  \pm  0.48 $  &   0.33  &   \\
   50719.84   & $   2.34  \pm  1.00 $  &   0.48  &   \\
   50720.85   & $   5.57  \pm  2.59 $  &   0.13  &   \\
   50721.83   & $   0.33  \pm  0.70 $  &   0.33  &   \\
\hline
\end{tabular}
\caption{\label{CT1 flux data} 
Diurnal flux values above 1.5 TeV as measured by CT1. For nights with
moon- and non-moon observations these fluxes are calculated as the weighted
mean. The Johnson V extinction coefficients (from the Carlsberg Automatic
Meridian Circle) are stated when available.}
\end{table}

\begin{table}
\begin{tabular}{|c|c|c|c|}
\hline
 Start & $ \Phi \left( \, {\rm E} > 1.5 \, {\rm TeV} \, \right) $ & Duration & Extinction \\
  $[\, \rm MJD \,]$ & $[\, 10^{-11} {\rm cm^{-2} s^{-1}}]\,$  & $[\, \rm h \,]$ & $[\,\rm Johnson V \,]$\\
\hline
50523.14 & $  1.74 \, \pm \,  0.38 $ &  2.43  & \\
50524.20 & $  2.49 \, \pm \,  0.49 $ &  1.17  & \\
50540.10 & $  2.14 \, \pm \,  0.98 $ &  0.65  & 0.13 \\
50545.18 & $  1.95 \, \pm \,  0.47 $ &  1.24  & \\
50546.08 & $  0.78 \, \pm \,  0.29 $ &  3.34  & 0.10 \\
50547.07 & $  1.28 \, \pm \,  0.32 $ &  3.50  & 0.11 \\
50548.07 & $  2.60 \, \pm \,  0.35 $ &  3.54  & \\
50549.07 & $  2.63 \, \pm \,  0.34 $ &  3.63  & \\
50550.10 & $  2.15 \, \pm \,  0.33 $ &  2.69  & 0.12 \\
50551.07 & $  7.08 \, \pm \,  0.57 $ &  2.47  & 0.12 \\
50552.06 & $  4.85 \, \pm \,  0.40 $ &  3.49  & 0.11 \\
50574.04 & $  1.97 \, \pm \,  0.47 $ &  1.50  & \\
50575.05 & $  3.32 \, \pm \,  0.66 $ &  1.00  & 0.10 \\
50577.05 & $  5.60 \, \pm \,  0.74 $ &  1.00  & 0.11 \\
50578.02 & $  4.91 \, \pm \,  0.58 $ &  1.50  & 0.13 \\
50599.97 & $  1.17 \, \pm \,  0.42 $ &  1.50  & \\
50600.97 & $  1.57 \, \pm \,  0.38 $ &  1.96  & 0.11 \\
50601.98 & $  2.46 \, \pm \,  0.40 $ &  2.00  & \\
50604.95 & $  4.22 \, \pm \,  0.48 $ &  1.99  & 0.12 \\
50606.97 & $  5.36 \, \pm \,  0.51 $ &  1.96  & 0.11 \\
50607.96 & $  2.95 \, \pm \,  0.60 $ &  1.00  & \\
50622.91 & $  5.79 \, \pm \,  0.84 $ &  0.83  & \\
50623.90 & $  3.71 \, \pm \,  0.49 $ &  1.67  & 0.11 \\
50624.91 & $  2.55 \, \pm \,  0.37 $ & 2.33 & 0.11 \\
50625.92 & $  7.54 \, \pm \,  0.53 $ & 2.26 & 0.10 \\
50626.91 & $  8.37 \, \pm \,  0.48 $ & 3.00 & \\
50627.95 & $  7.63 \, \pm \,  0.50 $ & 2.80 & 0.11 \\
50628.91 & $  1.29 \, \pm \,  0.29 $ & 2.83 & 0.11 \\
50629.91 & $  1.43 \, \pm \,  0.49 $ & 1.00 & \\
50630.90 & $  0.11 \, \pm \,  0.44 $ & 0.99 & \\
50631.91 & $  0.71 \, \pm \,  0.44 $ & 1.00 & 0.11 \\
50632.91 & $  0.39 \, \pm \,  0.44 $ & 0.99 & 0.14 \\
50633.91 & $  0.49 \, \pm \,  0.44 $ & 1.00 & 0.12 \\
50634.93 & $  1.81 \, \pm \,  0.75 $ & 0.50 & 0.12 \\
50635.90 & $  0.47 \, \pm \,  0.44 $ & 0.95 & \\
50636.91 & $  0.36 \, \pm \,  0.37 $ & 1.00 & 0.11 \\
50637.94 & $  0.67 \, \pm \,  0.38 $ & 1.00 & 0.20 \\
50638.96 & $  4.21 \, \pm \,  0.87 $ & 0.50 & 0.19 \\
50653.91 & $  2.39 \, \pm \,  0.59 $ & 0.81 & \\
50654.91 & $  1.23 \, \pm \,  0.65 $ & 0.49 & \\
50657.90 & $  1.73 \, \pm \,  0.66 $ & 0.50 & 0.14 \\
50658.90 & $  2.51 \, \pm \,  0.64 $ & 0.50 & 0.12 \\
50659.90 & $  1.50 \, \pm \,  0.61 $ & 0.50 & 0.11 \\
50660.90 & $  1.94 \, \pm \,  0.64 $ & 0.50 & 0.12 \\
50661.90 & $  2.04 \, \pm \,  0.68 $ & 0.50 & \\
50684.88 & $  1.98 \, \pm \,  0.44 $ & 1.76 & \\
50685.88 & $  1.61 \, \pm \,  0.46 $ & 1.50 & \\
50687.88 & $  1.47 \, \pm \,  0.55 $ & 1.00 & 0.10 \\
50688.88 & $ -0.05 \, \pm \,  0.40 $ & 1.49 & 0.10 \\
\hline
\end{tabular}
\caption{\label{CT2 flux data} 
Diurnal flux values above 1.5 TeV as measured by CT2. For some
nights we list the Johnson V extinction coefficients (from the Carlsberg
Automatic Meridian Circle).}
\end{table}

\section{The combined CT1, CT2 and CT system lightcurve above 1.5 TeV}
In Fig.~\ref{fig-lightcurve} the lightcurve from all HEGRA CTs is
shown for an energy  threshold of 1.5 TeV. The observations with
CT1 under the presence of moonlight are indicated separately. The errors are purely
statistical.

In general we see good agreement between the data from the three
instruments. Restricting the comparison to directly overlapping days we
obtain the following ratio between the fluxes
\begin{displaymath}
\frac{F ({\rm CT1})}{F ({\rm CT system})} = 0.73 \, \pm 0.02,
\ \ \ \frac{F ({\rm CT2})}{F ({\rm CT1})} = 1.03 \, \pm 0.04
\end{displaymath}
and
\begin{displaymath}
\frac{F ({\rm CT2})}{F ({\rm CT system})} = 0.89 \, \pm 0.03.
\end{displaymath}
The overlap times were 110 h, 60 h and 65 h, respectively. The seemingly
small inconsistency in the ratios has its origin in different overlap times
and different zenith angles. The ratios agree well within the systematic
errors, which are in the order of 30\%. The systematic error
is  in part global and in part quite
different for each instrument. There are 5 data points where the
disagreement between simultaneous observations with different telescopes
differ by more than 4 $\sigma$ (MJD 50579, 50580, 50582, 50607, 50658) 
after normalizing the fluxes to the respective mean fluxes.
These large differences are for the time being unexplainable. Part of the
difference might be due to source variability and different observation
times in compared nights but some of the discrepancies remain even for 
exactly matching ON time slices. 

Here we would like to comment on  various features of the lightcurve.
\begin{enumerate}
\item
The largest flare was observed at MJD 50626-27 with a flux above 1.5 TeV of
$\approx 10^{-10}$ cm$^{-2}$ sec$^{-1}$.
\item
Other experiments, Whipple and CAT, have observed a large and short flare
on April 16th (MJD 50554). Due to complete cloud coverage, HEGRA could not
observe that. 
\item
The deferred analysis of the data taken at large zenith angles will
add about 15\% more data points on the light curve as well as reducing
some of the errors of the shown data points.
\item
The data point at MJD 50526 is less reliable because observations were
carried out during strong and gusty winds.
\item
The visibility at MJD 50697 was at the allowed limit, therefore this flux
value may have to be corrected.
\item
A detailed time structure analysis including also the data from large
zenith angle observations (including observations under moonlight) is in
preparation and will be published elsewhere.
\item
No CT1/2 entries are shown after MJD 50721 because of modest statistics
below 38$^\circ$ zenith angle.
\end{enumerate}

\section{Discussion and summary}
\label{discussion}
The long period of intense flaring of Mkn 501 provided the unusual
opportunity
\begin{itemize}
\item[a)]
to obtain 
a large and relatively clean sample of VHE $\gamma$-rays,
\item[b)]
to carry out a detailed analysis of the spectral distribution
and the lightcurve over a duration of nearly seven months,
\item[c)]
to carry out multi-wavelength observations,
\item[d)]
to compare the highly variable $\gamma$ emission with that observed in 
previous years and also with that of other AGNs,
\item[e)]
to test the detector performance by comparing data taken at the same time
with nearby telescopes and in other experiments.
\end{itemize}

Most of the conclusions have already been presented in Part I. Here we
concentrate on a comparison of data taken with the different HEGRA CTs and
results related mainly to increased density of nightly samplings and
outline some future analysis prospects which will require additional
measurements.

The data of CT1 and 
2 presented here are for about half the time overlapping
with the CT system observation periods, while the other half fills many gaps,
dominantly during  moonlit nights, but normally CT1 was also
observing 30\%-50\%
longer during dark nights. Nevertheless, even during identical times and given 
the fact that CT1 is basically centered to the system, not only
identical events have been recorded. This
is due to the collection area of the CT system being about 2
1/2 times larger than that of CT1. Also, due to the larger cameras and
better precision on the impact parameter calculation, one could record with
the CT system showers where one ``sees'' only 
shower halo particles, i.e., from
showers with an impact distance between 130 - 200 m. Due to the different
readout concept, 8 bit FADC readout of the CT system and 10 respectively 11
bit charge sensitive gated ADCs of the stand-alone telescopes, 
the saturation
effects for multi-TeV showers are different. Also it should be
mentioned that we used different MC programs for the standalone
CT1/CT2 and the system. In addition we used quite
different procedures for the $\gamma$ selection. 
In spite of these differences we see in general excellent agreement
in the structure of the lightcurve
of the data recorded with the different instruments. In general, we see a
 better agreement between CT1 and CT2 data although their direct
event overlap is smaller than that between the CT system and CT1
observations. The flux values 
from CT1 are systematically lower compared to the CT system
data by about 27\%. 
For the time being we are unable to decide whether this is
related to the very different analysis methods or due to the systematic
errors in the photon to ADC signal conversion ratio.

\begin{figure}
\epsfxsize=8cm
\epsffile{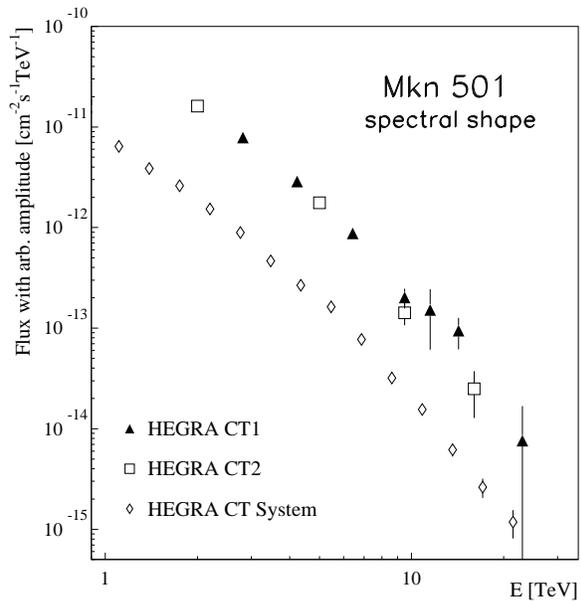}
\caption{\label{all_spec}
    The average spectral shape as measured by CT1, CT2 and the CT
system (data offset for clarity).}
\end{figure}

We observe good agreement of the spectral shape from the observations
with the different telescopes. Fig. \ref{all_spec} shows a comparative 
plot where we combined the two resp. three zenith angle ranges of CT1
and CT2. Obviously, an unbroken power law will not describe the data
well, except for the limited energy range of the CT1 data (see section 
\ref{sec-ct1spec}). An ansatz with an exponential cutoff
$$
        {\mathrm d}F/{\mathrm d}E \propto E^{-\alpha} \cdot e^{-E/E_0}
$$
yielded:\\
\\
for CT1:\ \ 
$
        \alpha = 2.09 \pm 0.09, \ E_0 = (7.16 \pm 1.04) \, \mathrm{TeV}
$\\
with a reduced\ $\chi^2 = 0.6$ (fit to the data points of
Fig. \ref{fig-ct1spec})\\
\\
for CT2:\ \ 
$
        \alpha = 1.27 \pm 0.37, \ E_0 = (2.85 \pm 0.58) \, \mathrm{TeV}
$\\
with a reduced\ $\chi^2 = 1.7$ (fit to the data points of
Fig. \ref{fig-ct2spec})\\
\\

The fit to the preliminary data of the HEGRA IACT system in the
energy region from 1.25 TeV to 24 TeV (Krawczynski \cite{krawc98}) gives:

for CT system:
$
        \alpha = 1.9 \pm 0.05, \ E_0 = (5.7 \pm 1.1) \, \mathrm{TeV}
$
\\

It should be noted that $\alpha$ and $E_0$ are highly
correlated, i.e., a modestly more curved spectrum enforces both a
lower $\alpha$ and lower E$_0$ in the fit. This is particularly
obvious for the CT2 data. The difference in the three sets of $\alpha$ 
and E$_0$ is explaineable by the different range of the fit.
If we fit the CT2 spectrum unsing the 'system' $\alpha$ of 
1.81 we obtain an $E_0 = (4.7 \pm 0.26)$ and a marginally worse
reduced\ $\chi^2$ of 1.71.
The observed steepening of the spectrum could be either due to an
inherent change in the acceleration and interaction process or due to
$\gamma$-interaction with the cosmic IR background. A detailed study
of the spectra will be presented in a forthcoming paper.

Other tests of the existence of an IR absorption have been proposed by
Aharonian (\cite{aharonian94}) and Plaga (\cite{plaga95}), 
namely the production of pair halos and
time delay of secondary $\gamma$s. The ``halo'' $\gamma$s should show up
predominantly at larger ALPHA values and should not show the rapid
variation of the main $\gamma$ flux at all. Another aspect is that, depending on
the onset of the IR absorption, the spectrum of the halo $\gamma$s should
be much softer resp. have a lower energy cutoff. The effect should be
quite visible in the 1-10 TeV region if strong IR absorption occurs above
25-35 TeV. We searched for such effects using data between 10$^\circ$ to
20$^\circ$ in ALPHA but no conclusive results could be drawn due to
insufficient statistics. One of the problems is that improvements in the
source position resolution due to higher energy can fake a soft halo
spectrum. Also the wide spread of the prediction for the extragalactic
magnetic field make such an analysis difficult.

Due to the large number of CT1 daily flux measurements a precise
comparison with the nearly continuous RXTE data is possible. The
correlation of 0.61 $\pm$ 0.06 between the CT1/2 data and the 
RXTE data  gives a rather strong evidence of a coupled
effect such as electron acceleration and inverse Compton scattering on
synchroton radiation generated photons. 

Close inspection of the data show that around MJD 50580 the structure of
the lightcurve in the TeV range differs significantly from that at the 2 -
10 keV range. While observing significant flaring in the TeV range, the keV
lightcurve remaines constant within errors. If we exclude the data
between MJD 50568 and 50590 the correlation rises to $0.65 \pm 0.07$
while inside the range it drops to $0.17 \pm 0.19$.

The assumption of the electron acceleration gets further
support from the about quadratic rise of TeV $\gamma$ flux as compared to
the keV flux rise from 1996 to 1997, see section \ref{sec-rxtecorr}. 
Clearly, a long term observation of
the keV - TeV correlation over a few years should give further support or
disprove the concept of electron acceleration dominance. Hadronic components
and/or a significant change in electron acceleration cannot be ruled out, see our
comment on the observation around MJD 50580. 


Note that the 1996 Mkn 501 data,
originally showing 
a 5.8 $\sigma$ excess (Bradbury et al. \cite{bradbury97}),
 have been reanalysed using
the dynamical supercuts also used in this paper. 
The excess increased to $> 7 \sigma$ while the
flux values and the integral spectrum remained the same 
($F(E > 1.5 {\mathrm{TeV}}) = 2.3 \times 10^{-12}\rm cm^{-2} s^{-1}$).
Comparing the 1996 and the 1997 spectrum, however, we find 
indications that the curvature of the spectrum has increased
since the increase in flux from 1996 to 1997 is lower at higher energies. 
A detailed comparison between the spectra in 1996 and 1997 will
be presented in a later paper. 

Finally, we want to comment on a technical conclusion drawn from the
analysis. For precision measurements it is important to use a larger camera
diameter compared to the one of CT1. A larger camera would have resulted in
a significantly better energy and angular resolution at higher energies.

\section*{Acknowledgements}

The HEGRA collaboration thanks the Instituto de Astrofisica de Canarias 
and the town of Garafia for 
use of the site and the excellent working conditions. 
Also we acknowledge the rapid availability of the RXTE data and the
atmospheric extinction data from the CAMC.
This work was supported
by the German Ministry of Education and Research, BMBF, 
the Deutsche Forschungsgemeinschaft, DFG,  and the Spanish Research
Foundation CICYT.

\end{document}